\documentclass{article}
\pdfoutput=1

\usepackage{amssymb}
\usepackage{color}
\usepackage[usenames,dvipsnames,svgnames,table]{xcolor}
\usepackage{soul}
\usepackage{setspace}

\def\be{\begin{equation}}
\def\ee{\end{equation}}
\def\ba{\begin{eqnarray}}
\def\ea{\end{eqnarray}}
\usepackage{mathptmx}
\usepackage{indentfirst}
\usepackage{amsmath,amssymb,latexsym,bm}
\usepackage{graphicx}	
\usepackage{amsmath}

\newcommand{\REV}[1]{\textcolor{blue}{#1}}

\usepackage{jcappub}

\begin{document}

\title{What Can We Learn by Combining  the Skew Spectrum and  the Power Spectrum?}
\author[a]{Ji-Ping Dai}

\author[b,c]{Licia Verde}

\author[a]{Jun-Qing Xia}

\affiliation[a]{Department of Astronomy, Beijing Normal University, Beijing 100875, China}
\affiliation[b]{Institut de Ciencies del Cosmos, University of Barcelona, ICCUB, Barcelona 08028, Spain.}
\affiliation[c]{Institucio Catalana de Recerca i Estudis Avancats, Passeig Lluis Companys 23, Barcelona 08010, Spain.}

\emailAdd{daijp@mail.bnu.edu.cn}
\emailAdd{liciaverde@icc.ub.edu}
\emailAdd{xiajq@bnu.edu.cn}

\abstract{
Clustering of the large scale structure provides complementary information to the measurements of the cosmic microwave background anisotropies through power spectrum and bispectrum of density perturbations. Extracting the bispectrum information, however, is more challenging than it is from the power spectrum due to the complex models and the computational cost to measure the signal and its covariance. To overcome these problems, we adopt a proxy statistic, skew spectrum which is a cross-spectrum of the density field and its quadratic field. By applying a large smoothing filter to the density field, we show the theory fits the simulations very well. With the spectra and their  full covariance estimated from  $N$-body simulations as our ``mock" Universe, we perform a global fits  for the cosmological parameters. The results show that  adding skew spectrum to power spectrum the $1\sigma$ marginalized errors for parameters  $ b_1^2A_s, n_s$ and $f_{\rm NL}^{\rm loc}$ are reduced by $31\%, 22\%, 44\%$, respectively. This { is the answer to the question posed in the title} and  indicates that the skew spectrum will be a fast and effective method  to access complementary information to that  enclosed in  the power spectrum measurements, especially for the forthcoming generation of wide-field galaxy surveys.}

\maketitle

\section{Introduction}
The origin of our Universe and its evolution have been  extensively probed by the cosmic microwave background (CMB)   anisotropies with a  three decades long effort that culminated with the Planck mission \citep{aghanim2018planck}. The next generation of CMB observations will also provide more precise measurements of the CMB polarization anisotropies \citep{abazajian2019cmb}. The large scale structure (LSS) of the Universe, that is, the distribution of matter and galaxies on large scales, is the result of the late-time evolution, powered by  gravitational instability,  of the same initial density perturbations responsible for the CMB anisotropies. Upcoming wide-field galaxy surveys, such as DESI \citep{aghamousa2016desi}, EUCLID \citep{amendola2018cosmology} and LSST \citep{Abell:2009aa}, are poised to provide  massive amount of high-precision data  carrying complementary information to that obtained  from the CMB measurements.

To date, most of the cosmological information from LSS is captured using 2-point clustering statistics, such as the 2-point correlation function or the power spectrum in Fourier space. However, the large scale structure we observe at low redshift is highly non-Gaussian as a result of the non-linear growth of structures, even for Gaussian initial conditions. Further  cosmological  information can be  obtained for the same surveys by using  also higher-order statistics, such as 3-point correlation function and bispectrum \citep{Matarrese:1997sk,Verde:1998zr, scoccimarro2000bispectrum, sefusatti2006cosmology, hoffmann2015measuring}. In particular the bispectrum  has been measured using galaxy survey data \citep{scoccimarro2001bispectrum, verde20022df, marin2013wigglez, gil2015power} and has proven useful to break degeneracies among cosmological parameters which arise from considering the power spectrum alone \cite{Gil-Marin:2014baa,Gil-Marin:2016wya}. Future LSS surveys  will enable  us to  reach a much larger signal-to-noise ratio for the bispectrum, providing a wealth of information, e.g., on primordial non-Gaussianity, non-linear bias and to further  reduce  the parameter degeneracies  present at the level of the power spectrum e.g.,~\cite{eggemeier2019bias} and Refs. therein.

However, extracting the information from the bispectrum is more challenging than it is from the power spectrum, due to the large number of triangle configurations and orientations.  Measuring the bispectrum signal and its covariance  requires  a significant computational effort  e.g., \citep{colavincenzo2018comparing}, and  the comparison of  theoretical models  with  measurements is rather complex e.g., \cite{Gil-Marin:2014baa,Gil-Marin:2016wya}.

In practice,  to bypass these challenges, several proxy statistics have been proposed to extract  (some of) the information  enclosed in the bispectrum; of particular interest are approaches that  compress the bispectrum to a pseudo-power spectrum, such that the signal depends only on one wavenumber (rather than three as  for the bispectrum). There are mainly two approaches: the integrated bispectrum proposed by Ref. \cite{chiang2014position} and the skew spectrum which was studied in CMB \citep{cooray2001squared, munshi2010new}, and then applied to LSS \citep{pratten2012non, schmittfull2015near, chan2017assessment, munshi2017integrated, dizgah2019capturing}, but see also pioneering works \cite{Bernardeau:1995ty, Pollack:2013alj}. The integrated bispectrum is generated by cross correlating the position-dependent power spectrum with the mean overdensity of the corresponding subvolume. This measurement contains parts of the bispectrum information on squeezed configuration; the application of  this statistic to real data  can be found in Ref. \citep{chiang2015position}. The skew spectrum (and later, the weighted skew spectrum)  are obtained by cross correlating the  (weighted) square of a field with the field itself. This quantity has recently received renewed attention \cite{schmittfull2015near, dizgah2019capturing}. Here we build on \cite{pratten2012non} and  partially also on these two works, with the aim of developing the approach further and  to bring it closer to a real application to  observations.  However, for simplicity,   here we do not consider the weighted skew spectrum.   In particular, we focus on prospects and possible challenges of  the joint analysis of skew spectrum and power spectrum, and compare theoretical and analytical modelling to $N$-body simulations.

As a first step, we  estimate the skew spectra from $N$-body simulations of the cosmological density field $\delta$ in a straightforward way by cross correlating $\delta(\boldsymbol x)$ and $\delta^2(\boldsymbol x)$, and then study the extra information that  can be obtained  by combining the  skew spectrum and the power spectrum. We consider three main sources of (non-Gaussian) skew spectrum signal: primordial non-Gaussianity,  gravitational instability and galaxy bias.
%\sout{We restrict ourselves to skew spectrum and power spectrum  in real space and omit the non-local bias term in this paper.
%\REV{Actually, we have checked our analysis by including the tidal shear bias and the redshift space distortions (RSD), and the conclusions did not change significantly.}}
The  full covariance matrix for the two statistics is estimated from a large suite of $N$-body simulations.  {This step, although challenging,  is particularly important, since  an accurate estimation  of the  full covariance of the two quantities is a key  ingredient to evaluate the added value offered by  a joint analysis of the two statistics.}  The  performance of our adopted {\it ansatz} for the likelihood function is also evaluated with a suite of simulations.
{All this enable us to quantify the usefulness of  using  skew spectrum and power spectrum jointly  to constrain some key cosmological parameters. This has not been addressed before in the literatures, but it  is a question worth asking before   deciding whether  to proceed to measure the skew spectrum from real galaxy surveys.  Note that  usually these type of analyses are done by Fisher matrix, which is much less computationally intensive. However, the high covariance between power spectrum and skew spectrum, and the complex properties of the skew-spectrum covariance,  force us to resort to numerically computed covariance from a suite of simulations. }

The rest of this paper is organized as follows. In section \ref{sec:met} we derive the full expression for the skew spectrum including primordial non-Gaussianity,  gravitational instability, galaxy bias {and  redshift space distortions}. We also compare the integrated bispectrum with the skew spectrum in this section. In section \ref{sec:sim} we present measurements of the skew spectra from $N$-body simulations and introduce the covariance used in our analysis. In section \ref{sec:result} we list the  resulting constraints using simulations and finally we conclude in section \ref{sec:con}.

\section{Methodology}
\label{sec:met}
{This section is somewhat pedagogical and  covers mostly background material, but is important to describe the methodology adopted and to define the notation used.}

Let us define the over density field $\delta({\boldsymbol x})=\delta\rho(\boldsymbol x)/\bar{\rho}$ where $\rho$ denotes the matter density field and $\bar{\rho}$ its (spatial) average. It is well known that the two point correlation function $\xi(r)=\langle \delta({\boldsymbol x})\delta({\boldsymbol x+\boldsymbol r})\rangle$, where $\langle . \rangle$ denotes the ensemble average is related to the power spectrum $P(k)$ via a Fourier transform. Similarly, the  3-point correlation function,
\be
\label{eq:3-cf}
\xi^{(3)}(\boldsymbol{x_{1}},\boldsymbol{x_{2}},\boldsymbol x_{3})=\left\langle\delta\left(\boldsymbol{x}_{1}\right) \delta\left(\boldsymbol{x}_{2}\right) \delta\left(\boldsymbol{x}_{3}\right)\right\rangle,
\ee
is related to the  bispectrum $B\left(\boldsymbol{k}_{1}, \boldsymbol{k}_{2}, \boldsymbol{k}_{3}\right)$ via,
\be
\xi^{(3)}\left(\boldsymbol{x}_{1}, \boldsymbol{x}_{2}, \boldsymbol{x}_{3}\right)=(2 \pi)^{3} \int_{\boldsymbol{k}_{1}} \int_{\boldsymbol{k}_{2}} \int_{\boldsymbol{k}_{3}} \delta^{D}\left(\boldsymbol{k}_{1}+\boldsymbol{k}_{2}+\boldsymbol{k}_{3}\right)
B_m\left(\boldsymbol{k}_{1}, \boldsymbol{k}_{2}, \boldsymbol{k}_{3}\right) {\rm e}^{\rm{i}\left[\boldsymbol{k}_{1} \cdot \boldsymbol{x}_{1}+\boldsymbol{k}_{2} \cdot \boldsymbol{x}_{2}+\boldsymbol{k}_{3} \cdot \boldsymbol{x}_{3}\right]},
\ee
where the Dirac delta $\delta^D$ ensures that the wavevectors correspond to the three sides of a triangle. The bispectrum is an effective statistic to recover information not present in power spectrum, but  it is challenging to measure form large scale structure data.

The (auto) skew spectrum is defined  from the cross correlation of the square of the field, $\delta^2$ with the  $\delta$ field  itself. { The squaring operation is intrinsically a configuration  space operation (very similar to local non-Gaussianity). For this reason  we find it natural to start from configuration space}.

%It is well known that the skew spectrum captures  part of the information present in the bispectrum.

In fact, let us  assume $\boldsymbol{x}_{3}$ in Eq.(\ref{eq:3-cf}) is  located  at the same point as $\boldsymbol{x}_{1}$:
\begin{equation}
\xi^{(3)}(\boldsymbol{x_{1}},\boldsymbol{x_{1}},\boldsymbol x_{2})= \xi^{(s)}(|\boldsymbol x_{1}-\boldsymbol x_{2}|)\equiv  \xi^{(s)}(x_{12})\,,
\end{equation}
where we have recognised the skew correlation function, $\xi^{(s)}$, and we have used the fact that the  cosmological principle imposes that  $\xi^{(s)}$ depends only on the  magnitude of the separation vector. The Fourier transform of  $\xi^{(s)}$ is the skew spectrum \citep{pratten2012non, schmittfull2015near, chan2017assessment, munshi2017integrated, dizgah2019capturing}.

We can now interpret the skew spectrum in light of the bispectrum. Following Ref. \cite{matarrese2008effect} we have\footnote{{ Ref. \cite{matarrese2008effect} deals with the effects of the bispectrum (induced by local primordial non-Gaussianity) on the power spectrum of thresholded objects. It is interesting to note the similarities here, thresholding like squaring are intrinsically configuration space operations, which then leave an effect on Fourier space  2-point quantities.}},
\be
\xi^{(s)}(x_{12} )
=\int \frac{d^{3} \boldsymbol{k}_{12}}{(2 \pi)^{3}} \int \frac{d^{3} \boldsymbol{k}_1}{(2 \pi)^{3}}
B_m(k_1,k_2,|\boldsymbol{k}_{12}|){\rm e}^{\rm{i}\boldsymbol{k}_{12}\boldsymbol{x}_{12}},
\ee
where $\boldsymbol x_{12} \equiv \boldsymbol{x}_1-\boldsymbol{x}_2$ and $\boldsymbol{k}_{12} \equiv \boldsymbol{k}_1+\boldsymbol{k}_2$. The Fourier transform of this function yields  the skew spectrum which can be written as
\be
\label{eq:gen}
P_m^{(s)}(k)=\int \frac{d^{3} \boldsymbol{q}}{(2 \pi)^{3}}B_m(k,q,|\boldsymbol{q}-\boldsymbol{k}|)
=\int_{-1}^1 d\mu \int \frac{d{q}}{(2 \pi)^2}q^2B_m(k,q,\alpha(\mu)),
\ee
we have adopted the replacement: $\boldsymbol{k}_{12}\rightarrow \boldsymbol{k}, \boldsymbol{k}_{1}\rightarrow\boldsymbol{q}, \boldsymbol{k}_{2}\rightarrow\boldsymbol \alpha$
where $\mu=\boldsymbol{k}\cdot\boldsymbol{q}/kq$, and $\alpha=\sqrt{q^2+k^2-2\mu k q}$.
The skew spectrum encloses information beyond Gaussianity. Now we consider three main sources of non-Gaussianity in real space: primordial non-Gaussianity, non-Gaussianity  from  gravitational instability and non-Gaussianity from galaxy bias. We  begin by discussing  these three effects  separately, {see also Ref. \citep{pratten2012non, schmittfull2015near, dizgah2019capturing}}.

\subsection{Primordial Non-Gaussianity}
The skew spectrum on large scales is  sensitive to  the statistical properties of  the primordial fluctuations. For example, non-Gaussianity of the local type is given by \citep{salopek1990nonlinear, Gangui:1993tt, verde2001tests, komatsu2001acoustic},
\be
\Phi(\boldsymbol{x})=\Phi_{G}(\boldsymbol{x})+f_{\rm NL}^{\rm{loc}} \left[\Phi_{G}^{2}(\boldsymbol{x})-\left\langle\Phi_{G}^{2}(\boldsymbol{x})\right\rangle\right],
\ee
where $\Phi(\boldsymbol{x})$ denotes the Bardeen's curvature perturbation during the matter era, $\Phi_{G}(\boldsymbol{x})$ is a Gaussian field and $f_{\rm NL}^{\rm{loc}}$ is a constant which characterizes the amplitude of primordial non-Gaussianity. The leading contribution to the bispectrum of the curvature field is given by
\be
B_{\Phi}\simeq 2f_{\rm NL}^{\rm{loc}}\left[ P_{\Phi}(k_1)P_{\Phi}(k_2)+\rm cyc. \right],
\ee
where $P_{\Phi}(k)=\left\langle\Phi(k)\Phi^{*}(k)\right\rangle$. For the local type non-Gaussianity, most of the signal is concentrated in  the  so-called squeezed triangular configurations, $k_1\ll k_2, k_3$.

Density fluctuations in Fourier space, $\delta(k)$, are related to the curvature perturbations,
\be
\delta(k)=M(k,a)\Phi(k);~ M(k,a) =\frac{2k^2 T(k) D(a)}{3\Omega_{m}H_{0}^{2}},
\ee
where $a$ is the scale factor, $H_0$ is the current Hubble constant, $\Omega_m$ is the current  matter energy density parameter, $T(k)$ is the matter transfer function and $D(a)$ is the growth factor. This allows us to write the contribution to the primordial matter bispectrum as
\be
B_{m,I}(k_1,k_2,k_3)=M(k_1)M(k_2)M(k_3)B_{\Phi}(k_1,k_2,k_3)\,
\ee
here we have omitted $a$ for brevity. The matter skew spectrum caused by the primordial non-Gaussianity is
\be
\label{eq:pris}
P_{m,I}^{(s)}(k)= 2f_{\rm NL}^{\rm loc}M(k)P_\Phi(k)\int_{-1}^1 d\mu \int \frac{d{q}}{(2 \pi)^2}q^2 M(q)P_\Phi(q)M(\alpha)
\left[ 2+\frac{P_\Phi(\alpha)}{P_\Phi(k)} \right].
\ee

There are other non-Gaussian templates,  motivated by general single-field models of inflation, yielding  bispectra such as equilateral model $B_{\Phi}^{\rm eq}=6 f_{\rm NL}^{\rm eq}F^{\rm eq}$ \citep{seery2005primordial, chen2007observational} and orthogonal model $B_{\Phi}^{\rm or}=6 f_{\rm NL}^{\rm or}F^{\rm or}$ \citep{senatore2010non}, where
\be
F^{\rm eq}\simeq-\left(P_{\Phi}\left(k_{1}\right) P_{\Phi}\left(k_{2}\right)+2 \rm {cyc.}\right)-2\left[P_{\Phi}\left(k_{1}\right) P_{\Phi}\left(k_{2}\right) P_{\Phi}\left(k_{3}\right)\right]^{2 / 3}+\left(P_{\Phi}\left(k_{1}\right)^{1 / 3} P_{\Phi}\left(k_{2}\right)^{2 / 3} P_{\Phi}\left(k_{3}\right)+5 \rm {cyc.}\right),
\ee
\be
F^{\rm or}\simeq-3\left(P_{\Phi}\left(k_{1}\right) P_{\Phi}\left(k_{2}\right)+2 \rm {cyc.}\right)-8\left[P_{\Phi}\left(k_{1}\right) P_{\Phi}\left(k_{2}\right) P_{\Phi}\left(k_{3}\right)\right]^{2 / 3}+3\left(P_{\Phi}\left(k_{1}\right)^{1 / 3} P_{\Phi}\left(k_{2}\right)^{2 / 3} P_{\Phi}\left(k_{3}\right)+5 \rm {cyc.}\right).
\ee
We stress here that these are templates, their correspondence to explicit  non-Gaussian models, especially in the limit of specific configurations,  is not perfect. Nevertheless, as it is widespread in the literature,  we work here with these templates which we sometimes refer to as {\it shapes}.
The skew spectra for these non-Gaussian shapes are obtained simply as,
\be
P_{m,I}^{(s)}(k)^{\rm eq(or)}=M(k)\int_{-1}^1 d\mu \int \frac{d{q}}{(2 \pi)^2}q^2M(q)M(\alpha) B^{\rm eq(or)}_{\Phi}(k,q,\alpha).
\ee

\subsection{Non-Gaussianity from gravitational instability}
Even for  Gaussian initial conditions, the late-time non-linear gravitational evolution generates a non-zero bispectrum. At quasi-linear scales  non-linear evolution of matter density fluctuations  can be modelled by perturbation theory in which case the density field  is expanded as [e.g., \citep{bernardeau2002large}]
\be
\delta({\boldsymbol{k}})=\delta({\boldsymbol{k}})^{(1)}+\delta({\boldsymbol{k}})^{(2)}+\delta({\boldsymbol{k}})^{(3)}+\ldots,
\ee
here we truncate expansions at the second order, and $\delta({\boldsymbol{k}})^{(2)}$ is given by,
\be
\delta({\boldsymbol{k}})^{(2)}=\int \mathrm{d}^{3} \boldsymbol q_{1} \mathrm{d}^{3} \boldsymbol q_{2} \delta_{D}\left(\boldsymbol{k}-\boldsymbol{q}_{12}\right) F_{2}\left(\boldsymbol{q}_{1}, \boldsymbol{q}_{2}\right) \delta({\boldsymbol{q}_{1}})^{(1)} \delta({\boldsymbol{q}_{2}})^{(1)},
\ee
where $F_2(\boldsymbol{q}_{1}, \boldsymbol{q}_{2})$ is the known second-order kernel of standard perturbation theory,
\be
F_{2}\left(\boldsymbol{q}_{1}, \boldsymbol{q}_{2}\right)=\frac{5}{7}+\frac{x}{2}\left(\frac{q_{1}}{q_{2}}+\frac{q_{2}}{q_{1}}\right)+\frac{2}{7} x^{2},
\label{eq:2OPTkernel}
\ee
with $x \equiv {\boldsymbol{q}}_{1} \cdot {\boldsymbol{q}}_{2}/q_1q_2$.

At leading order, the gravitational instability bispectrum is
\be
B_{m,G}\left(k_{1}, k_{2}, k_{3}\right)=2 F_{2}\left(\boldsymbol{k}_{1}, \boldsymbol{k}_{2}\right) P_{m,L}\left(k_{1}\right) P_{m,L}\left(k_{2}\right)+\mathrm{cyc.},
\ee
where $P_{m,L}\left(k\right)$ is the linear matter power spectrum. Hereafter, we use the subscript ``$L$'' to represent the linear terms. This expression for the perturbative bispectrum can of course be improved. A particularly interesting
modification is the phenomenological one proposed by Ref. \cite{scoccimarro2001fitting, gil2012improved}, which maintains the same structure and adjusts the coefficients of Eq.(\ref{eq:2OPTkernel}) to fit $N$-body simulations.  The matter skew spectrum contribution from non-linear gravitational evolution is therefore,
\be
P_{m,G}^{(s)}(k)=\int_{-1}^1 d\mu \int \frac{d{q}}{(2 \pi)^2}q^2B_{m,G}(k,q,\alpha).
\ee

\subsection{Non-Gaussianity from galaxy bias}
Halos and galaxies are biased tracers of the dark matter field. In our analysis, we use a simple prescription in Eulerian space, where the galaxy overdensity is expanded in terms of the matter overdensity and the traceless part of the tidal tensor. Up to quadratic order, we have [e.g., \citep{desjacques2018large}]
\be
\delta_{g}(\boldsymbol{x}) \simeq b_{1} \delta(\boldsymbol{x})+\frac{1}{2} b_{2} \delta^{2}(\boldsymbol{x})+
\frac{1}{2}b_{K^2}\left[\left(\frac{\partial_{i} \partial_{j}}{\partial^{2}}-\frac{1}{3} \delta_{i j}\right)\delta(\boldsymbol{x})\right]^2~,
\ee
{where $b_1, b_2$ represent the linear and non-linear bias and $b_{K^2}$ describes the non-local tidal shear bias. In Fourier space this becomes,
\be
\delta_{g}(\boldsymbol{k}) \simeq b_1 \delta(\boldsymbol{k})+\frac{1}{2}b_2\int \mathrm{d}^{3} {\boldsymbol q} \delta({\boldsymbol q})\delta{({\boldsymbol k-\boldsymbol q})} + \frac{1}{2}b_{K^2}\int \mathrm{d}^{3} {\boldsymbol q} \delta({\boldsymbol q})\delta{({\boldsymbol k-\boldsymbol q})} S_2(\boldsymbol q,\boldsymbol k-\boldsymbol q)~,
\ee
where $S_2$ is defined from the Fourier transform of the tidal tensor
\be
S_2(\boldsymbol{q}_{1}, \boldsymbol{q}_{2}) = \frac{(\boldsymbol{q}_{1} \cdot {\boldsymbol{q}}_{2})^2}{(q_1 q_2)^2}-\frac{1}{3}~.
\ee
}

{Since we are only interested in exploring the complementarity between power spectrum and skew spectrum and the relative reduction on the size of posterior errors of key cosmological parameters,  we are not overly concerned  about adopting the latest bias model, as long as the modelling adopted is a good description of the simulations (see Sec.~\ref{sec:sim}).}
For simplicity, we assume that the galaxy (or halo) formation is a local process. %depends only on the local matter density field, so we only keep the first two terms.
 {A physically motivated choice for $b_{K^2}$  (also adopted e.g., in the bispectrum analysis of BOSS data \cite{Gil-Marin:2014baa}) is  to consider  a local bias  in Lagrangian space \cite{Baldauf:2012hs, Chan:2012jj}, where  $b_{K^2}=-4/7(b_1-1)$. A careful inspection of Fig.~9 of \cite{schmittfull2015near}  however, indicates that for the halo skew spectrum, the sensitivity to  $b_{K^2}$ is very small, we will return to this below, but we anticipate that setting $b_{K^2}=0$ does not have any significant effect in this work.}
%
%In principle, rather than assuming  $b_{K^2}=0$ In fact  since we are only interested in exploring the complementarity between power spectrum and skew spectrum and the relative reduction on the size of posterior errors of key parameters. Actually, We have checked the case by including the tidal bias term, and we find it does not influence our final results significantly.}

%Non-local bias can be
%studied with the cross skew spectrums between density field, displacement field and tidal field as Ref. \cite{schmittfull2015near}  pointed out. A more detailed treatment of galaxy bias is left for future work, our simplified
%model shall suffice as at this level.

 The galaxy bispectrum with primordial part and gravitational part becomes,
%\be
%B_g\left(k_{1}, k_{2}, k_{3}\right)=b_{1}^{3} \left[ B_{m,I}\left(k_{1}, k_{2}, k_{3}\right)+B_{m,G}\left(k_{1}, k_{2}, k_{3}\right)\right]+b_{1}^2b_{2}\left[P_{m,L}\left(k_{1}\right) P_{m,L}\left(k_{2}\right)+\rm{cyc.}\right]{+B_{K^2}\left(k_{1}, k_{2}, k_{3}\right)}.
%\ee
%{where  $B_{K^2}\left(k_{1}, k_{2}, k_{3}\right)=2b_{K^2}S_2(k_1,k_2) P_{m,L}\left(k_{1}\right) P_{m,L}\left(k_{2}\right)+\mathrm{cyc.}$}.

{\be
\begin{aligned}
B_g\left(k_{1}, k_{2}, k_{3}\right)=&b_{1}^{3} \left[ B_{m,I}\left(k_{1}, k_{2}, k_{3}\right)+B_{m,G}\left(k_{1}, k_{2}, k_{3}\right)\right]\\
&+b_{1}^2\left[b_2P_{m,L}\left(k_{1}\right) P_{m,L}\left(k_{2}\right)+b_{K^2}P_{m,L}\left(k_{1}\right) P_{m,L}\left(k_{2}\right)S_2(\boldsymbol{k_1},\boldsymbol{k_2})+\rm{cyc.}\right].
\end{aligned}
\ee}

\subsection{Full expression for skew spectrum of biased  tracers}

The galaxy (or halo) skew spectrum can thus be written factoring out  the  galaxy power spectrum (the equations are shown for local type non-Gaussianity).
\be
\label{eq:pg}
P_g^{(s)}(k)=\mathcal{F}(k)P_{g,L}(k),
\ee
\be
\label{eq:mathf}
\mathcal{F}(k)=\int_{-1}^1 d\mu \int \frac{d{q}}{(2 \pi)^2}q^2\left[ C_1 P_{g,L}(q)+C_2 P_{g,L}(\alpha)+C_3 \frac{P_{g,L}(q)P_{g,L}(\alpha)}{P_{g,L}(k)}\right],
\ee
where
\be
C_1=\frac{1}{b_{1}} \left[2f_{\rm NL}^{\rm loc}\frac{M(\alpha)}{M(k)M(q)}\left( 2+\frac{P_\Phi(\alpha)}{P_\Phi(k)} \right)+2{F_2\left(\boldsymbol{k},\boldsymbol{q}\right)}+ \frac{b_2}{b_1}{+\frac{b_{K^2}}{b_1}S_2\left(\boldsymbol{k},\boldsymbol{q}\right)}\right], \\
\ee
\be
 C_2=\frac{1}{b_{1}}\left[ 2{F_2\left(\boldsymbol{k},\boldsymbol{\alpha}\right)} + \frac{b_2}{b_1}  {+\frac{b_{K^2}}{b_1}S_2\left(\boldsymbol{k},\boldsymbol{\alpha}\right)}\right],\\
\ee
\be
 C_3=\frac{1}{b_{1}}\left[ 2{F_2\left(\boldsymbol{q},\boldsymbol{\alpha}\right)} + \frac{b_2}{b_1} {+\frac{b_{K^2}}{b_1}S_2\left(\boldsymbol{q},\boldsymbol{\alpha}\right)}\right].\\
\ee
%{\be
%C_1=\frac{1}{b_{1}} \left[2f_{\rm NL}^{\rm loc}\frac{M(\alpha)}{M(k)M(q)}\left( 2+\frac{P_\Phi(\alpha)}{P_\Phi(k)} \right)+2{F_2\left(\boldsymbol{k},\boldsymbol{q}\right)}+ \frac{b_2}{b_1}+\frac{b_{K^2}}{b_1}S_2(\boldsymbol{k},\boldsymbol{q})\right], \\
%\ee
%\be
% C_2=\frac{1}{b_{1}}\left[ 2{F_2\left(\boldsymbol{k},\boldsymbol{\alpha}\right)} + \frac{b_2}{b_1}+\frac{b_{K^2}}{b_1}S_2(\boldsymbol{k},\boldsymbol{\alpha})\right],\\
%\ee
%\be
% C_3=\frac{1}{b_{1}}\left[ 2{F_2\left(\boldsymbol{q},\boldsymbol{\alpha}\right)} + \frac{b_2}{b_1}+\frac{b_{K^2}}{b_1}S_2(\boldsymbol{q},\boldsymbol{\alpha})\right].\\
%\ee}

\begin{figure*}[tb]
	\centering
    \includegraphics[width=0.49\linewidth]{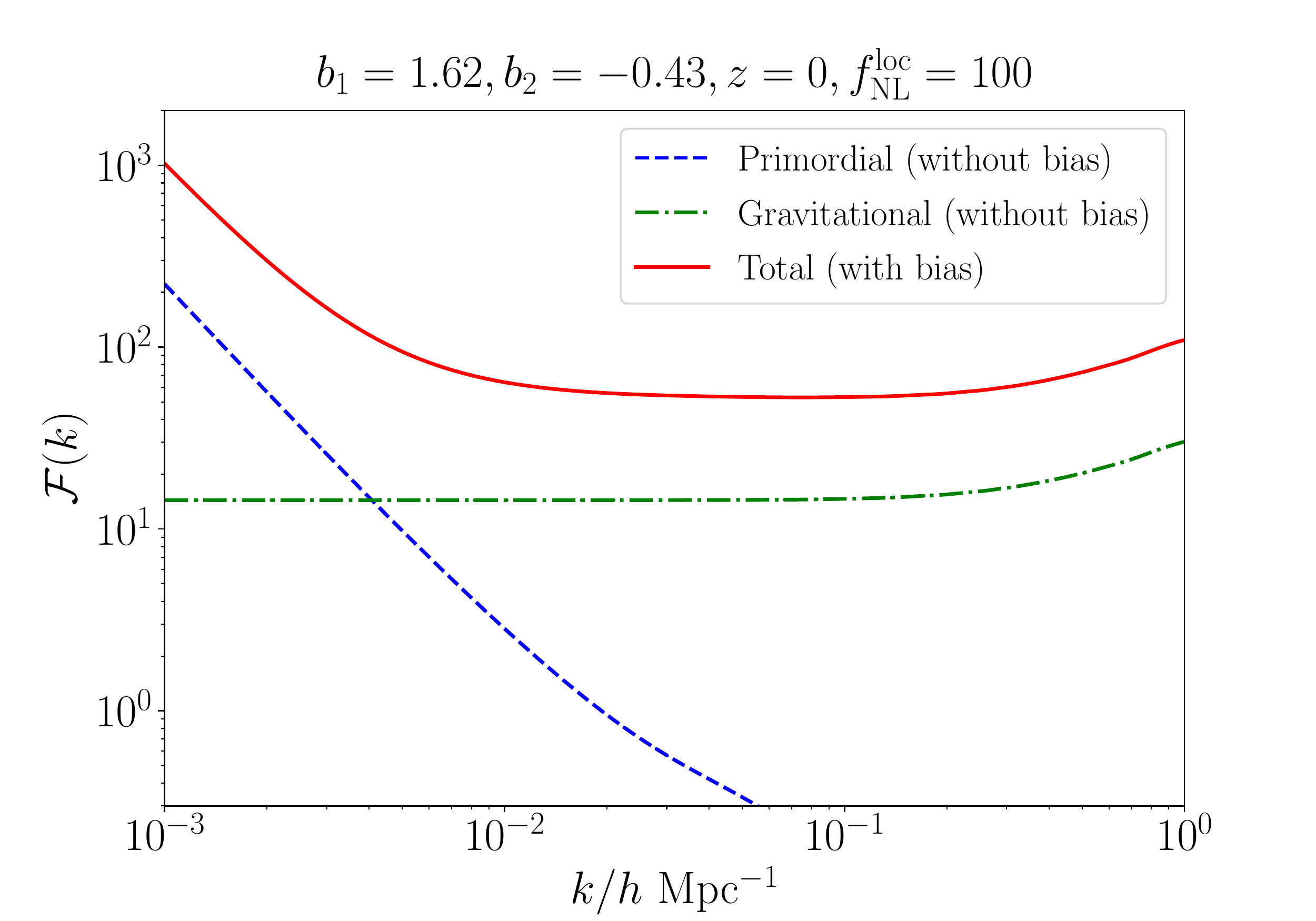}
    \includegraphics[width=0.49\linewidth]{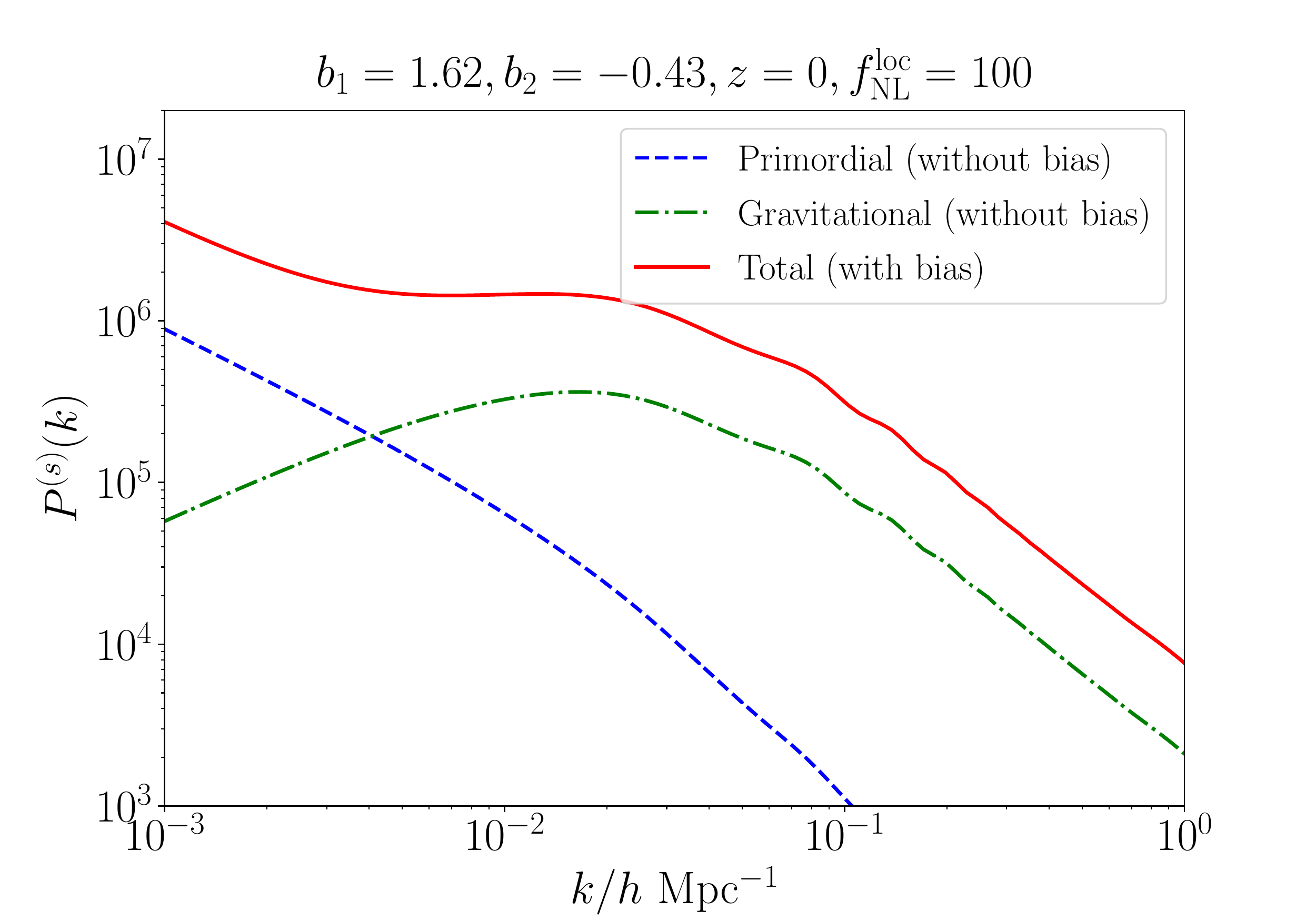}
    \includegraphics[width=0.49\linewidth]{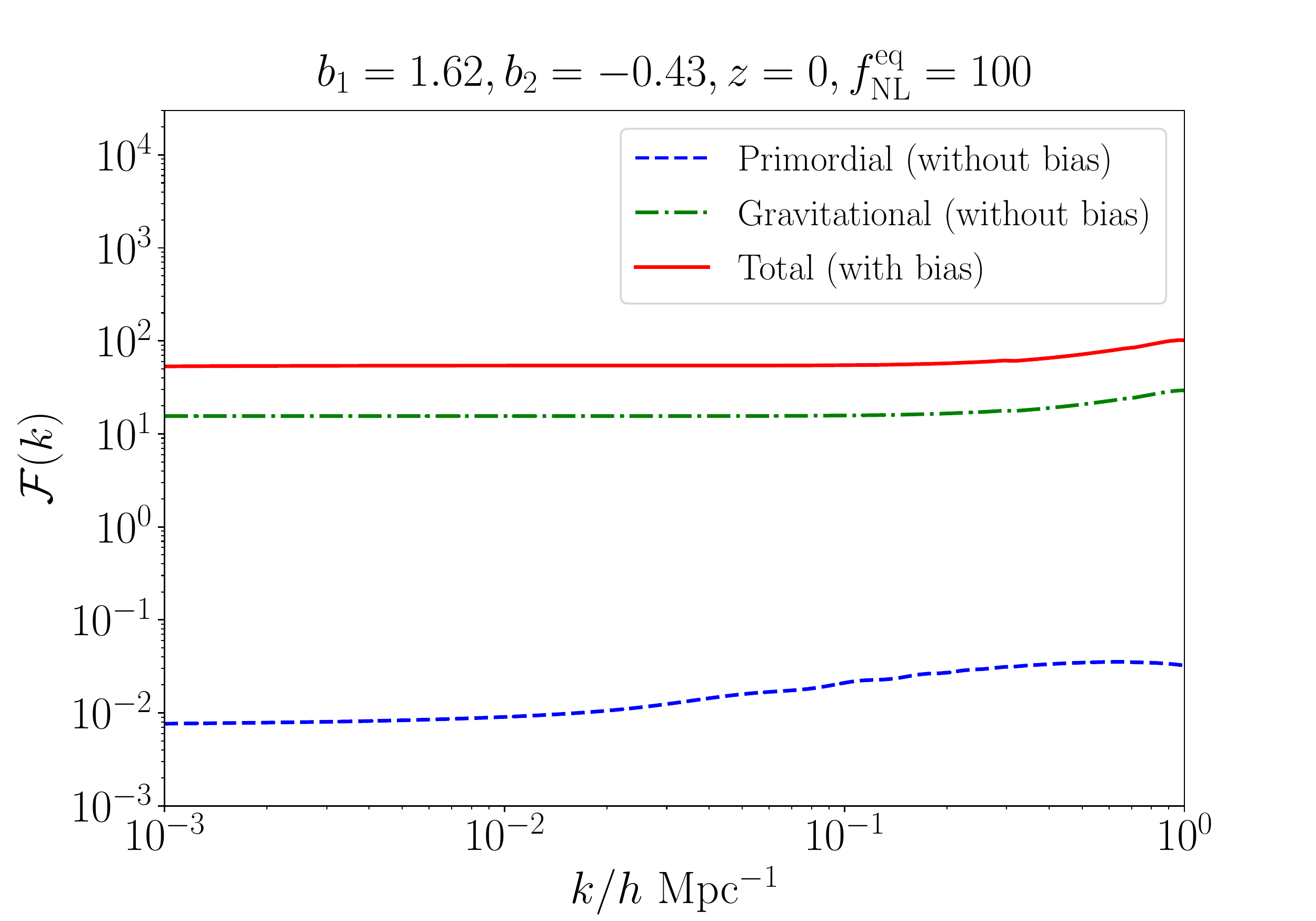}
    \includegraphics[width=0.49\linewidth]{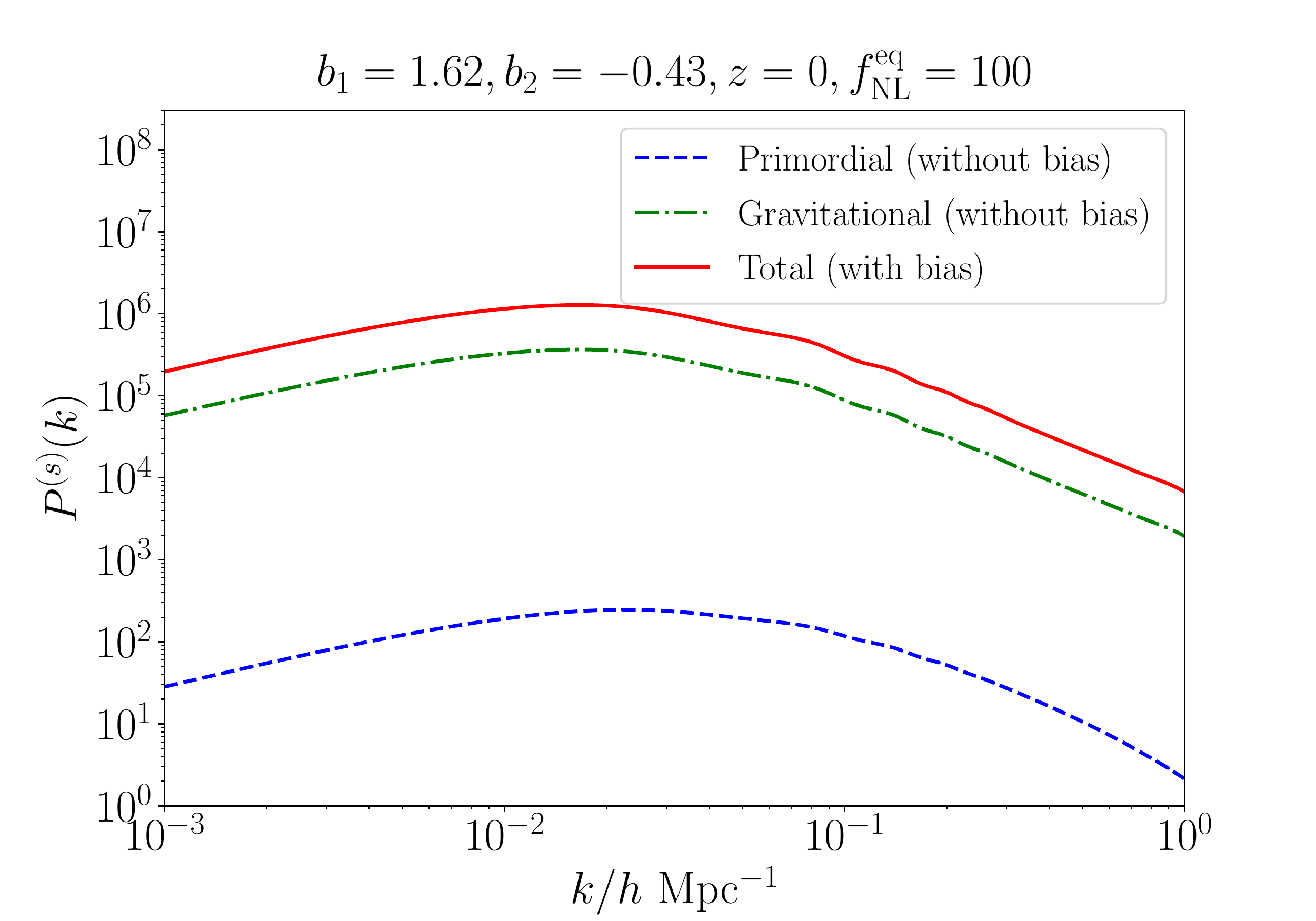}
    \includegraphics[width=0.49\linewidth]{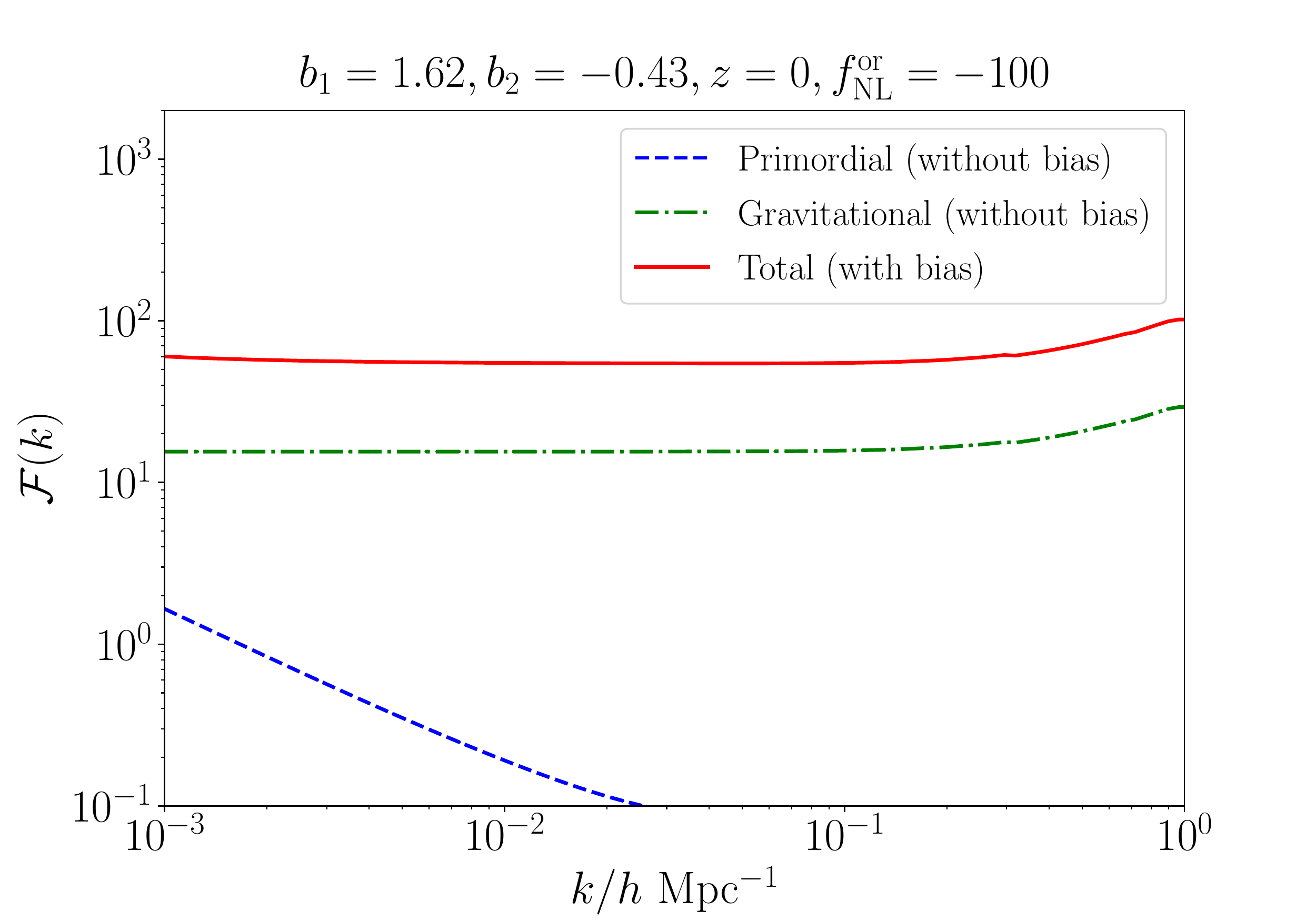}
    \includegraphics[width=0.49\linewidth]{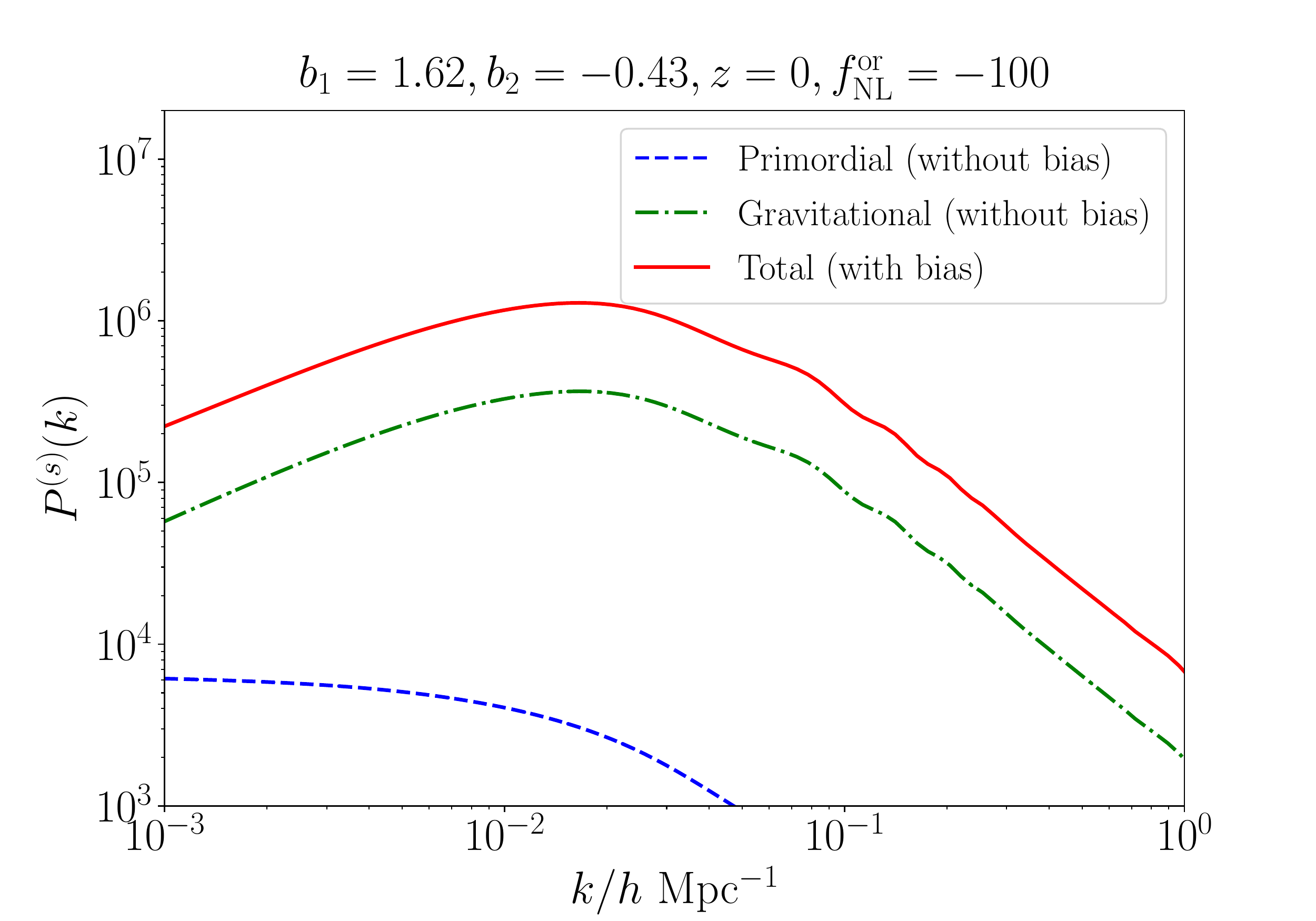}
	\caption{Left panel: The function  of $\mathcal{F}(k)$ in Eq.(\ref{eq:pg}) , for primordial non-Gaussianity for unbiased tracers at $z=0$ ($f^{\rm loc}_{\rm NL} = f^{\rm eq}_{\rm NL} =100, f^{\rm or}_{\rm NL}=-100$, blue dashed line), gravitational instability also for unbiased tracers at $z=0$ (green dotted-dashed line) and their combination for biased tracers  (red solid line) with bias parameters $b_1=1.62, b_2=-0.43$,  {$b_{K^2}=0$}; Right panel: Same as the left panel but for the  skew spectra. The linear power spectrum adopted is that of a Planck LCDM cosmology. We also show the results for three different non-Gaussianity templates (upper: local; middle: equilateral; lower: orthogonal.)}
	\label{fig:skew}
\end{figure*}

Here $\mathcal{F}(k)$ depends weakly on cosmological parameters (via the weak dependence of the perturbation theory kernel and the transfer function), depends explicitly on  bias parameters and depends linearly on the non-Gaussianity parameter $f^{\rm loc}_{\rm NL}$, which  indicates  how  the skew spectrum  carries  information  additional to that encoded in the the power spectrum.

The left panels of Fig. \ref{fig:skew} shows the function of $\mathcal{F}(k)$ in Eq.~(\ref{eq:pg}) for three different non-Gaussianity templates (local, equilateral and orthogonal), gravitational instability, and their combination for biases tracers with: $f_{\rm NL}^{\rm loc}=f_{\rm NL}^{\rm eq}=100, f_{\rm NL}^{\rm or}=-100$, $z=0$ $b_1=1.62, b_2=-0.43$, {$b_{K^2}=0$}. This bias corresponds to that of halos above a minimum mass  $M_{\min}=2.5\times10^{13}h^{-1}M_{\odot}$ at $z=0$ (see Sec. \ref{sec:result}).

For gravitational instability, $\mathcal{F}(k)$ is almost a constant on linear scales.
This is not unexpected.  To understand this in a simple way let us  consider unbiased tracers. In this case,  the function of $\mathcal{F}(k)$ for gravitational instability is,
\be
\mathcal{F}(k)=2\int_{-1}^1 d\mu \int \frac{d{q}}{(2 \pi)^2}q^2\left[ F_2(\boldsymbol{k},\boldsymbol{q}) P_{g,L}(q)+F_2(\boldsymbol{k},\boldsymbol{\alpha}) P_{g,L}(\alpha)+F_2(\boldsymbol{q},\boldsymbol{\alpha}) \frac{P_{g,L}(q)P_{g,L}(\alpha)}{P_{g,L}(k)}\right],
\ee
if $\boldsymbol{k} \rightarrow 0$, $\boldsymbol{q}\simeq- \boldsymbol{\alpha}$,
so $F_2(\boldsymbol{q},\boldsymbol{\alpha})=0$. We can simplify the function as,
\ba
\mathcal{F}(k)|_{k\rightarrow0}&=&4\int_{-1}^1 d\mu \int \frac{d{q}}{(2 \pi)^2}q^2 F_2(\boldsymbol{k},\boldsymbol{q}) P_{g,L}(q)\\
&=&4\int_{-1}^1 d\mu \int \frac{d{q}}{(2 \pi)^2}q^2\left( \frac{5}{7}+\frac{2}{7} \mu^{2}\right) P_{g,L}(q)+4\int_{-1}^1 d\mu \int \frac{d{q}}{(2 \pi)^2}q^2\left[ \frac{\mu}{2}\left( \frac{k}{q}+\frac{q}{k} \right) \right] P_{g,L}(q).
\ea
The first term is independent of $k$ and the second term  goes  to 0 because $\int_{-1}^1 d\mu C(k)\mu = 0$. Then it is proved that for gravitational instability, $\mathcal{F}(k)$ is almost a constant on large scales.

Primordial non-Gaussianity affects  mostly large scales, $k<0.03~h\rm Mpc^{-1}$.  The  corresponding skew spectra (primordial non-Gaussianity,  gravitational instability for unbiased tracers and their combination for biased tracers) are shown in the right panel of Fig. \ref{fig:skew}, also with $f_{\rm NL}^{\rm loc}=f_{\rm NL}^{\rm eq}=100, f_{\rm NL}^{\rm or}=-100, z=0, b_1=1.62, b_2=-0.43$, \REV{$b_{K^2}=0$}. We find the local type non-Gaussianity has the most prominent effect at large scales, and the gravitational part becomes important at smaller scales.
For the same value of  the $f_{\rm NL}$ parameter, the other two templates only have a much weaker impact on the final skew spectra. %\textcolor[rgb]{1.00,0.00,0.00}{\textbf{Maybe should be reworded}: Interestingly, the ratio of $\mathcal{F}(k)$ for equilateral (orthogonal) template primordial non-Gaussianity to  that for local primordial non-Gaussianity is roughly constant on large scales, although this approximation holds much better for the equilateral case than the local one. This can be appreciated in Fig.~\ref{fig:ratioFng}.}

\subsubsection{Comparison with other approaches to access the bispectrum information via the power spectrum}
Another approach proposed to access bispectrum information via a suitable power spectrum is the ``integrated bispectrum" proposed by Ref. \cite{chiang2014position}, which measures an integral of the bispectrum  which is dominated by the squeezed configurations.

This statistics is obtained by dividing the survey volume $V$ into $N_s$ subvolumes.
In each sub volume, the local power spectrum (the so-called position-dependent power spectrum, $P(\boldsymbol k,\boldsymbol {r_L})$)  and  local mean over-density , $\bar \delta_{r_L}$, are computed. Then correlating the position-dependent power spectrum with the local mean over density one obtains the so-called integrated bispectrum.

\be
\begin{aligned}\left\langle P\left(\boldsymbol{k}, \boldsymbol{r}_{L}\right) \bar{\delta}\left(\boldsymbol{r}_{L}\right)\right\rangle=&\frac{1}{V_{L}^{2}} \int \frac{d^{3} \boldsymbol q_{1}}{(2 \pi)^{3}} \int \frac{d^{3} \boldsymbol q_{3}}{(2 \pi)^{3}} B_m\left(\boldsymbol{k}-\boldsymbol{q}_{1},-\boldsymbol{k}+\boldsymbol{q}_{1}+\boldsymbol{q}_{3},-\mathbf{q}_{3}\right) \\ &\times W_{L}\left(\boldsymbol{q}_{1}\right) W_{L}\left(-\boldsymbol{q}_{1}-\boldsymbol{q}_{3}\right) W_{L}\left(\boldsymbol{q}_{3}\right), \end{aligned}
\ee
where $V_L = V/N_s$, $\boldsymbol{r}_{L}$ is the center of the subvolume and $W_L$ are the window functions.

Because of the window functions, most of the contribution to the integrated bispectrum thus comes from values of $q_1$ and $q_3$ until approximately $1/\sqrt[3]{V_L}$. Ref. \cite{chiang2014position} pointed out that, if the wavenumber $\boldsymbol k$ is much larger than $1/\sqrt[3]{V_L}$, then the dominant contribution to the integrated bispectrum comes from the bispectrum in squeezed configurations, $B\left(\boldsymbol{k}-\boldsymbol{q}_{1},-\boldsymbol{k}+\boldsymbol{q}_{1}+\boldsymbol{q}_{3},-\boldsymbol{q}_{3}\right) \rightarrow B\left(\boldsymbol{k},-\boldsymbol{k},-\boldsymbol{q}_{3}\right)$ with $q_1\ll k$ and $q_3\ll k$. The integrated bispectrum becomes,
\be
\left\langle P\left(\boldsymbol{k}, \boldsymbol{r}_{L}\right) \bar{\delta}\left(\boldsymbol{r}_{L}\right)\right\rangle\simeq \frac{1}{V_{L}^{2}} \int^{1/\sqrt[3]{V_L}} \frac{d^{3} \boldsymbol q_{3}}{(2 \pi)^{3}} B\left({k},{k},{q}_{3}\right).
\label{eq:2.30}
\ee
By comparing this expression with the  skew spectrum
\be
P^{(s)}(k)=\int \frac{d^{3} \boldsymbol{q}}{(2 \pi)^{3}}B(k,q,|\boldsymbol{q}-\boldsymbol{k}|),
\label{eq:2.31}
\ee
we can appreciate that the two quantities are highly complementary; {the integrated bispectrum is mostly  sensitive to squeezed configurations while the skew spectrum is sensitive to a combination of all shapes. However, in the limit  where $q_3 <1/\sqrt[3]{V_L}$ and $k\gg q_3$ (i.e., $q_3 \longrightarrow 0$)  in  Eq.~(\ref{eq:2.30})  the skew spectrum is equivalent to the  integrated bispectrum proposed by Ref. \cite{chiang2014position} for  $k\longrightarrow 0$ in Eq.~(\ref{eq:2.31}).}

%$q<1/\sqrt[3]{V_L}$ and $k\gg q$, the skew spectrum is equivalent to the  integrated bispectrum proposed by Ref. \cite{chiang2014position}.

\subsection{Smoothing}
When comparing our  theoretical predictions for the skew spectrum  to data or $N$-body simulations, we should consider that the evolved field is likely highly non-linear. Our derived expression is valid on quasi-linear scales, and is expected to fail in the non-linear regime. There are several fitting formulae for the dark matter bispectrum we can use to derive a more reliable expression for the skew spectrum \citep{scoccimarro2001fitting, gil2012improved}. However these formulae are only calibrated (and valid) in a specific  $k$ range. To avoid this problem, we apply a smoothing filter to the field  to suppress the small scales non-linear  modes.

\begin{figure}[htb]
	\centering
    \includegraphics[width=0.8\linewidth]{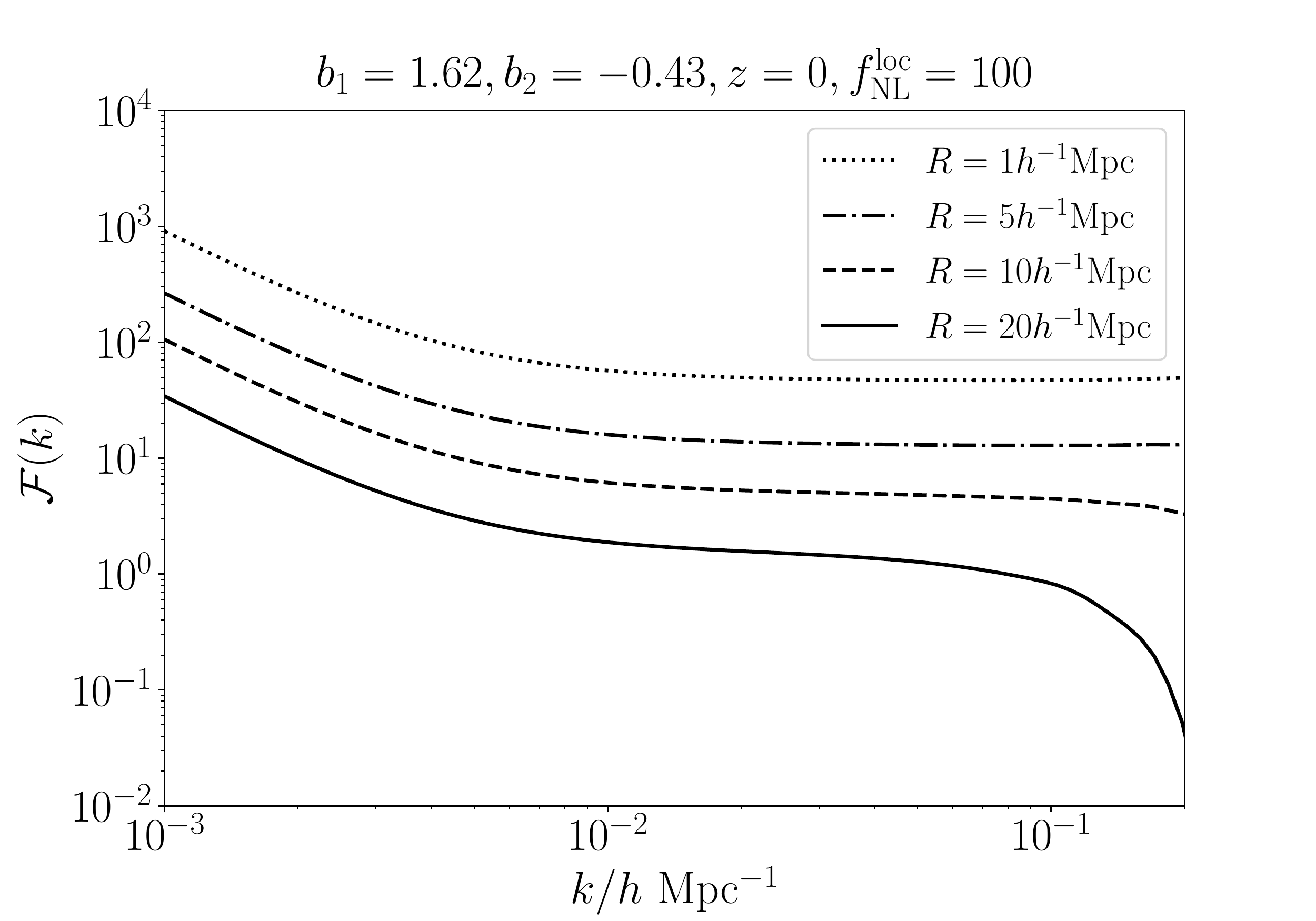}
	\caption{$\mathcal{F}(k)$ for different smoothing radii, from top to bottom: $R=1, 5, 10, 20 h^{-1}\rm Mpc$}
	\label{fig:Fsm}
\end{figure}

By doing this, we may loose some information from the skew spectrum, but we can have analytical control.
In this paper, we use a top-hat windows function whose Fourier transform is,
\be
W_{R}(k)=\frac{3 \sin (k R)}{k^{3} R^{3}}-\frac{3 \cos (k R)}{k^{2} R^{2}}.
\ee
The smoothed skew spectrum becomes
\ba
P_R^{(s)}(k) &=& \int \frac{d^{3} \boldsymbol{q}}{(2 \pi)^{3}}B(k,q,\alpha)W_R(k)W_R(q)W_R(\alpha).
\ea
In Fig. \ref{fig:Fsm} we show ${\cal F}(k)$ with $f_{\rm NL}^{\rm loc}=100, z=0, b_1=1.62, b_2=-0.43$, for different smoothing radii. The introduction of a smoothing filter reduces the amplitude of ${\cal F}(k)$ and therefore also of $P^{(s)}(k)$. In Sec. \ref{sec:sim} we compare the analytic expression for the skew spectrum to the  measurements from $N$-body simulations.

{
\subsection{Redshift space distortions}
In this section, we extent our theory by including the RSD effect which is caused by the peculiar velocities of galaxies in the redshift measurements of surveys. This distortion depends on the growth rate of structures,  and therefore in principle  it can offer complementary information.}

{
At tree-level, the galaxy bispectrum with local primordial non-Gaussianity can be conveniently written as \cite{Scoccimarro:1999ed, gil2014dark, Tellarini:2016sgp}
\be
B^{\rm RSD}_g(k_1,k_2,k_3)=2 Z_{2}\left(\boldsymbol{k}_{1}, \boldsymbol{k}_{2}\right) Z_{1}\left(\boldsymbol{k}_{1}\right) Z_{1}\left(\boldsymbol{k}_{2}\right) P_{m,L}\left(k_{1}\right) P_{m,L}\left(k_{2}\right)+\mathrm{cyc.}
\ee
where the kernels $Z_i$ are defined as
\be
Z_1(\boldsymbol{k}_{i})\equiv(b_1+f\nu_i^2)
\ee
\ba
Z_2(\boldsymbol{k}_{1},\boldsymbol{k}_{2})&\equiv&b_{1}\left[F_{2}\left(\boldsymbol{k}_{1}, \boldsymbol{k}_{2}\right)+\frac{f \nu k}{2}\left(\frac{\nu_{1}}{k_{1}}+\frac{\nu_{2}}{k_{2}}\right)+f^{\rm loc}_{\rm NL}\frac{M(k_3)}{M(k_1)M(k_2)}\right]+  \nonumber\\
&&f \nu^{2} \left[G_{2}\left(\boldsymbol{k}_{1}, \boldsymbol{k}_{2}\right)+f^{\rm loc}_{\rm NL}\frac{M(k_3)}{M(k_1)M(k_2)}\right]+\frac{f^{2} \nu k}{2} \nu_{1} \nu_{2}\left(\frac{\nu_{2}}{k_{1}}+\frac{\nu_{1}}{k_{2}}\right)+\frac{b_{2}}{2}~,
\ea
where $f$ is the logarithmic growth rate ${\rm d} \ln \delta /{\rm d} \ln a$, $\nu_i$ denote the cosine of the angle between $\boldsymbol{k}_i$ and the line of sight,
$\nu\equiv\left(\nu_{1} k_{1}+\nu_{2} k_{2}\right) / k$ and  $k^{2}=\left(\boldsymbol{k}_{1}+\boldsymbol{k}_{2}\right)^{2}$. $G_2(\boldsymbol{k}_{1},\boldsymbol{k}_{2})$ is the second-order kernels of the velocities,
\be
G_{2}\left(\boldsymbol{k}_{1}, \boldsymbol{k}_{2}\right)=\frac{3}{7}+\frac{x}{2}\left(\frac{k_{1}}{k_{2}}+\frac{k_{2}}{k_{1}}\right)+\frac{4}{7} x^{2}
\ee
with $x \equiv {\boldsymbol{k}}_{1} \cdot {\boldsymbol{k}}_{2}/k_1k_2$.
}
{While  the anisotropic signal of redshift space distortions for the power spectrum have been studied extensively, for the bispectrum  only the angle-averaged (monopole) signal has  been measured (see \cite{Gualdi:2020ymf}).}
{A closed expression for the bispectrum monopole can only  be obtained  for large scales where the non-linear  Fingers-of-God (FoG) effect is small. This is indeed the case in the regime we are interested in.}
{ Here we model the  bispectrum monopole as
%To investigate the angle averaged reshot space  bispectrum (bispectrum monopole), we can decompose the redshift space bispectrum in Legendre basis and the halo bispectrum monopole can be analytically written as
\ba
B^{(0)}_g(k_1,k_2,k_3)&=&b_1^4\left\{\frac{1}{b_{1}} \left[F_{2}\left(\boldsymbol{k}_{1}, \boldsymbol{k}_{2}\right)+f^{\rm loc}_{\rm NL}\frac{M(k_3)}{M(k_1)M(k_2)} \right] \mathcal{D}_{\mathrm{SQ} 1}^{(0)} + \frac{1}{b_{1}} \left[G_{2}\left(\boldsymbol{k}_{1}, \boldsymbol{k}_{2}\right)+f^{\rm loc}_{\rm NL}\frac{M(k_3)}{M(k_1)M(k_2)} \right] \mathcal{D}_{\mathrm{SQ} 2}^{(0)} \right. \nonumber\\
&&\left. +\frac{b_2}{b_1^2}\mathcal{D}_{\rm NLB}^{(0)}+ \mathcal{D}_{\rm FoG}^{(0)}\right\}P_{m,L}(k_1)P_{m,L}(k_2)+\rm cyc.
\ea
where the terms $\mathcal{D}_{\mathrm{SQ} 1}^{(0)}$ and $\mathcal{D}_{\mathrm{SQ} 2}^{(0)}$ represent the linear and non-linear contributions to the large-scale squashing, $\mathcal{D}_{\rm NLB}^{(0)}$ is due to the non-linear biasing and $\mathcal{D}_{\rm FoG}^{(0)}$ describes (partially) the effect of damping due the Fingers of God. These terms are defined in \cite{Scoccimarro:1999ed}.
}

{
Finally, we can write the smoothed galaxy skew spectrum monopole in redshift space as
\be
P_g^{(s,0)}(k) = \int \frac{d^{3} \boldsymbol{q}}{(2 \pi)^{3}}B_g^{(0)}(k,q,\alpha)W_R(k)W_R(q)W_R(\alpha).
\ee
}
{ The monopole  redshift space  power spectrum is described by
\be
P_g^{0}(k)=\left[1+\frac{2}{3}\frac{f}{b_1}+\frac{1}{5}\left(\frac{f}{b_1}\right)^2\right]P_{g,L}(k)
\ee
 where  we  also neglect the non-linear Fingers of God effects.}

{ Of course it would not make sense to consider the  real-world complications due to redshift space distortions, if the answer to the question posed in the title was "nothing" or ``very little" even in real space. For this reason we will first  present   real-space results and then generalise them to redshift space. This generalisation comes with a caveat: here we only consider the monopole signal (and large, linear scales), adding higher order multipoles such as the quadrupole might change  quantitatively the results. But this is left for future work.}

\section{Simulations}
\label{sec:sim}
We use 1000 realizations  from the $\textsc{Quijote}$ simulations suite \footnote{https://github.com/franciscovillaescusa/Quijote-simulations} \citep{villaescusa2019quijote}.  The simulation's  cosmological parameters are $\Omega_{\rm m}=0.3175, \Omega_{\rm b}=0.049, h=0.6711, n_s=0.9624, \sigma_8=0.834, M_\nu=0.0~{\rm eV}$, and $f_{\rm NL}^{\rm loc}=0$, which are the matter and baryon density parameters, reduced Hubble constant, spectral index of primordial power law power spectrum, amplitude of perturbations parameter, total neutrino mass and amplitude of primordial non-Gaussianity respectively. All the simulations were run using the TreePM code $\textsc{Gadget-III}$, an improved version of $\textsc{Gadget-II}$ \citep{springel2005cosmological}. Each realization has $512^3$ cold dark matter particles in a box with cosmological volume of 1$(h^{-1}\rm Gpc)^3$. We use the halo catalogues where halos were identified using the Friends-of-Friends algorithm \citep{davis1985evolution} with linking length $b=0.2$ at $z=0$, and we set the minimal halo mass $M_{\min}=2.5\times10^{13}h^{-1}M_{\odot}$. Details of the simulations can be found in \citep{villaescusa2019quijote}.

{These simulations have Gaussian initial conditions. Below we will explore  forecasted error-bars also for local non-Gaussianity parameter $f^{\rm loc}_{\rm NL}$.  Since the simulations have  $f^{\rm loc}_{\rm NL}=0$ the resulting errors should be considered valid for a {\it null-test  hypothesis}. In case of a significant detection of non-zero $f^{\rm loc}_{\rm NL}$ error-bars are expected to be different.}

To calculate the skew spectra from simulations, we square the (smoothed) halo density field and treat it as a new field $\delta^2_h(\boldsymbol x)$, then calculate the cross power spectrum with $\delta_h(\boldsymbol x)$ using the routine provided in $\textsc{Pylians}$ \footnote{https://github.com/franciscovillaescusa/Pylians}. When considering smoothing, we apply a top-hat smoothing filter with $R=20h^{-1}\rm Mpc$ before squaring the density field.  {With this smoothing choice we find that standard perturbation theory is sufficient to describe the skew spectra. This smoothing filter also ensures that in redshift space,  non-linear Fingers-of-God effects are   negligible.}

\begin{figure}[htb]
	\centering
    \includegraphics[width=0.8\linewidth]{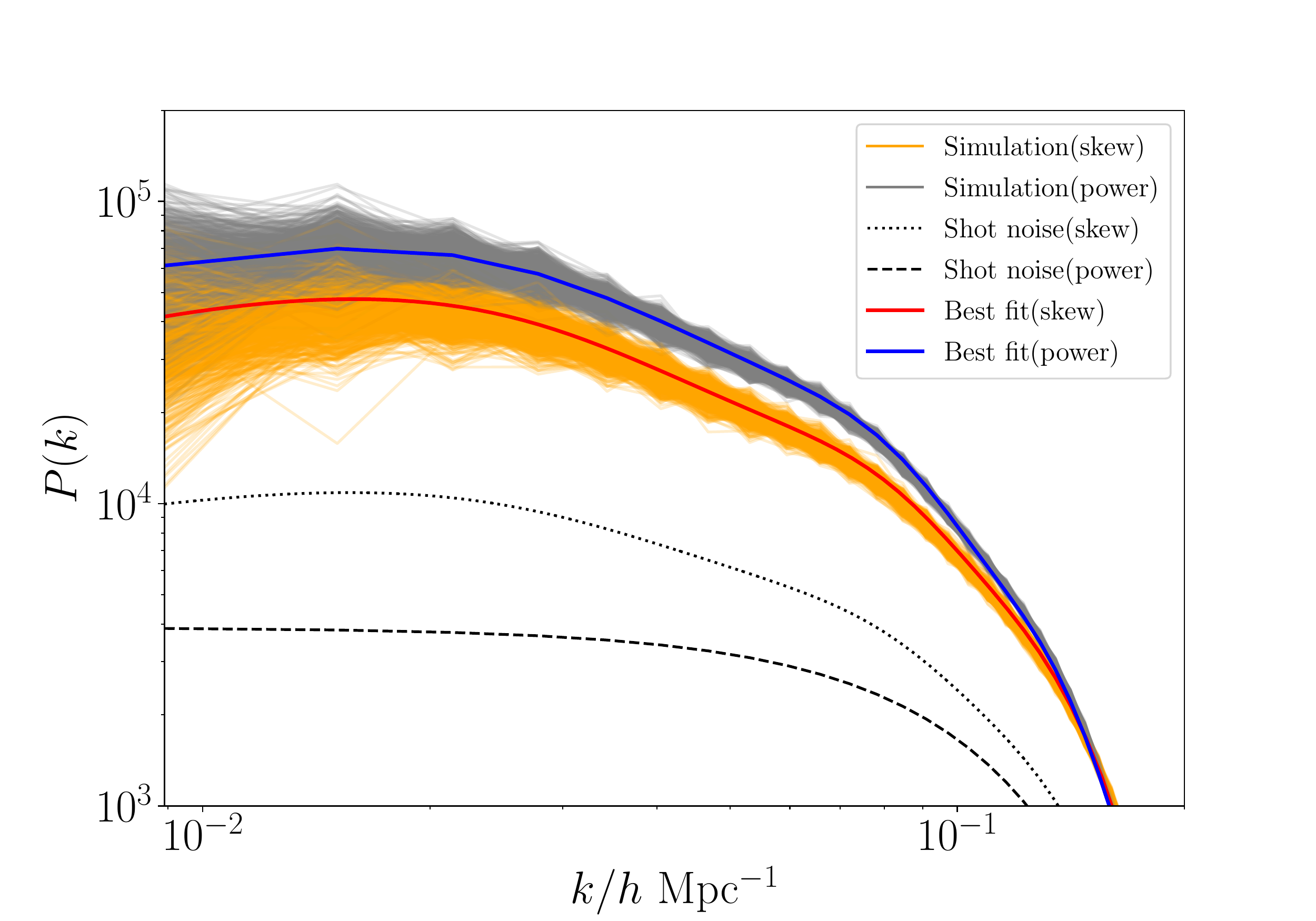}
	\caption{Power spectra (grey lines) and skew spectra (orange lines) from 1000 simulations, the black dotted line and dashed line are shot noise for skew spectrum and power spectrum, respectively.  Solid lines correspond to the best-fit theoretical models (details can be found in Sec. \ref{sec:result}). The smoothing radius is $R=20 h^{-1}\rm Mpc$}
	\label{fig:shot}
\end{figure}

\subsection{Shot noise}
\label{sec:shot}
Both power spectrum and skew spectrum of the halo density have an additional stochasticity contribution (shot noise) whose Poissonian predictions are \citep{schmittfull2015near}
\be
P_{h, {\rm measured}}(k)=P_{h}(k)+ P_{h,\rm shot}(k) \,;\,\,\, P_{h,\rm shot}(k)= \frac{1}{n_h},
\ee
and for the skew spectrum,
\be
P^{(s)}_{h,\rm shot}(k) = \int \frac{d^{3} \boldsymbol{q}}{(2 \pi)^{3}} \left[\frac{1}{n_h}\left(P_h(k)+ P_h(q)+P_h(|\boldsymbol{q}-\boldsymbol{k}|) \right)+\frac{1}{n_h^2}\right],
\ee
where $n_h$ is the halo number density and $P_h(k)$ is the halo power spectrum with shot noise subtracted. Because of halo-exclusion e.g., \citep{casas2002distribution, bonatto2011proper}, we cannot expect the shot noise contribution to be exactly Poissonian. Following Ref. \cite{gil2014dark}, who used a simple, 1-free parameter model for the halo shot noise, and  found the shot noise  for a halo population to  be slight sub-Poisson, we adopt the following parameterizations,

\be
P_{h,\rm shot}(k)=(1-A_{\rm noise})\frac{1}{n_h},
\label{eq:SNpk}
\ee
\be
P^{(s)}_{h,\rm shot}(k)=(1-A_{\rm noise})\int \frac{d^{3} \boldsymbol{q}}{(2 \pi)^{3}} \left[\frac{1}{n_h}\left( P_h(k)+ P_h(q)+P_h(|\boldsymbol{q}-\boldsymbol{k}|)-\frac{3A_{\rm noise}}{n_h} \right)+\frac{1}{n_h^2}\right],
\label{eq:snPskew}
\ee
here $A_{\rm noise}$ is a free parameter to account for deviations from Poisson behaviour.
{Neither from BOSS data \cite{Gil-Marin:2014baa} nor from the simulations we use here there is evidence for  the $A_{\rm noise}$ phenomenological shot noise correction to  take different  values  for the power spectrum  and  the bispectrum  or for the   the power spectrum (Eq.~(\ref{eq:SNpk})) and the skew spectrum  (Eq.~(\ref{eq:snPskew})). Here we therefore adopt only one  $A_{\rm noise}$ parameter for both statistics. While this assumption may not hold  in detail for future data, it is not expected to  change the conclusions of this work.} {Since in practice a  good prior on $A_{\rm noise}$ may be obtained   e.g., from scale smaller than those considered here. We present our main results by  fixing  $A_{\rm noise}$ in the joint analysis to its best fit value obtained for a fixed (underling) cosmology. We then also show how results change when  $A_{\rm noise}$ is allowed to float  along with  all the other parameters. We anticipate here that the  first approach is slightly more conservative: including  the skew spectrum  give in this case  slightly smaller gains (see also appendix~\ref{appendix:Anoise}).}

In Fig.~\ref{fig:shot}
we plot the  real-space power spectra (gray) and skew spectra (orange) obtained from the $z=0$  smoothed halo field (for halos above $M_{\min}=2.5\times 10^{13} h^{-1}M_{\odot}$) of the simulations and their corresponding shot noise.  Each of the thin lines correspond to one simulation. The thick blue and red lines are our theoretical predictions for the best-fit parameters in particular  for bias and shot noise which are $A_{\rm noise}=0.22, b_1=1.635, b_2=-0.426$.  These values of the linear bias and of $A_{\rm noise}$ coincide with those obtained by fitting the ratio of the power spectra of the halos and  of the dark matter at scales $k<0.1$ $h$/Mpc (see details in Sec. \ref{sec:result}).

The skew spectra from simulations in Fig. \ref{fig:shot} are smaller than the theory prediction (right panel of Fig. \ref{fig:skew}), because here  we use a smoothing filter and the contribution from smaller scales is suppressed.

{The corresponding plots in redshift space (monopole) are qualitatively very similar  reported in appendix \ref{appendix:RSD}.}
%\LV{can you include the same plots in redshift space here? report the values of $\beta$ and $\sigma_v$ adopted.}

\subsection{Covariances}

Before being able to perform a  joint analysis of power spectrum and skew spectrum, we need to
evaluate the full covariance of both of these two quantities. We estimate it from the $\textsc{Quijote}$ simulations using a wavenumber range  $k=[0.0089, 0.1] h\rm Mpc^{-1}$,  in 15 $k$ bins uniformly  spaced in log $k$.
We start by combining  $P_h(k)$ and $P_h^{(s)}(k)$ into a ``data" vector $P^{(p+s)}_h(K_i)$ ($i=1,\ldots,15$ for the  power spectrum and $i=16,\ldots,30$ for the skew spectrum).

\begin{figure}[htb]
	\centering
    \includegraphics[width=0.43\linewidth]{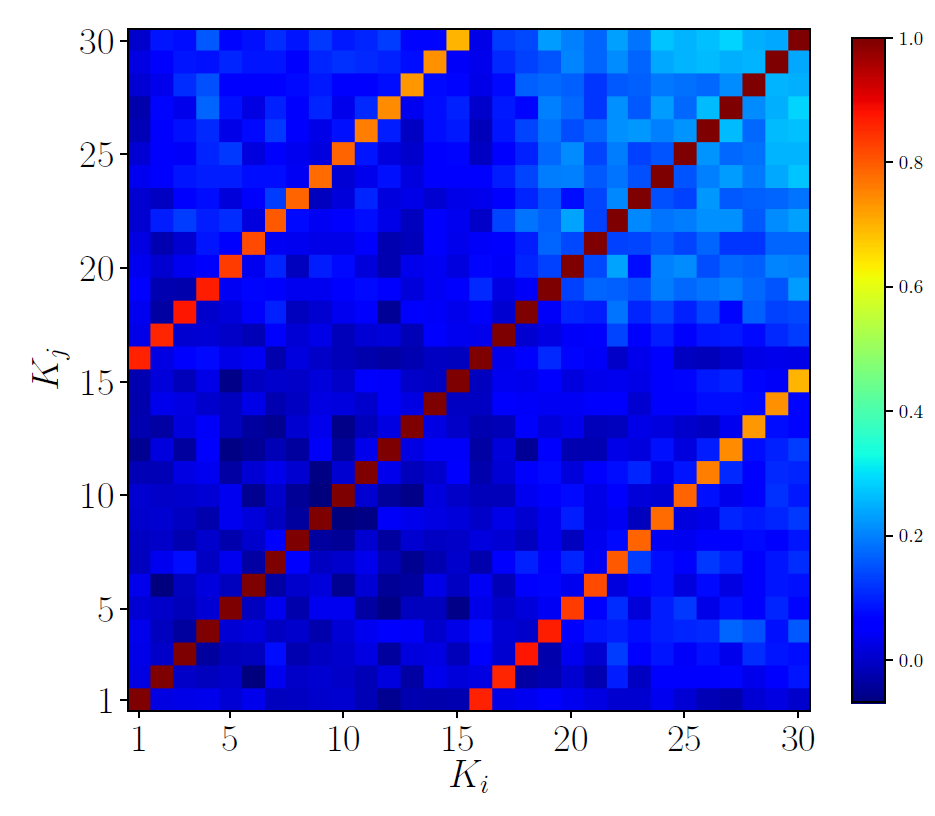}
    \includegraphics[width=0.56\linewidth]{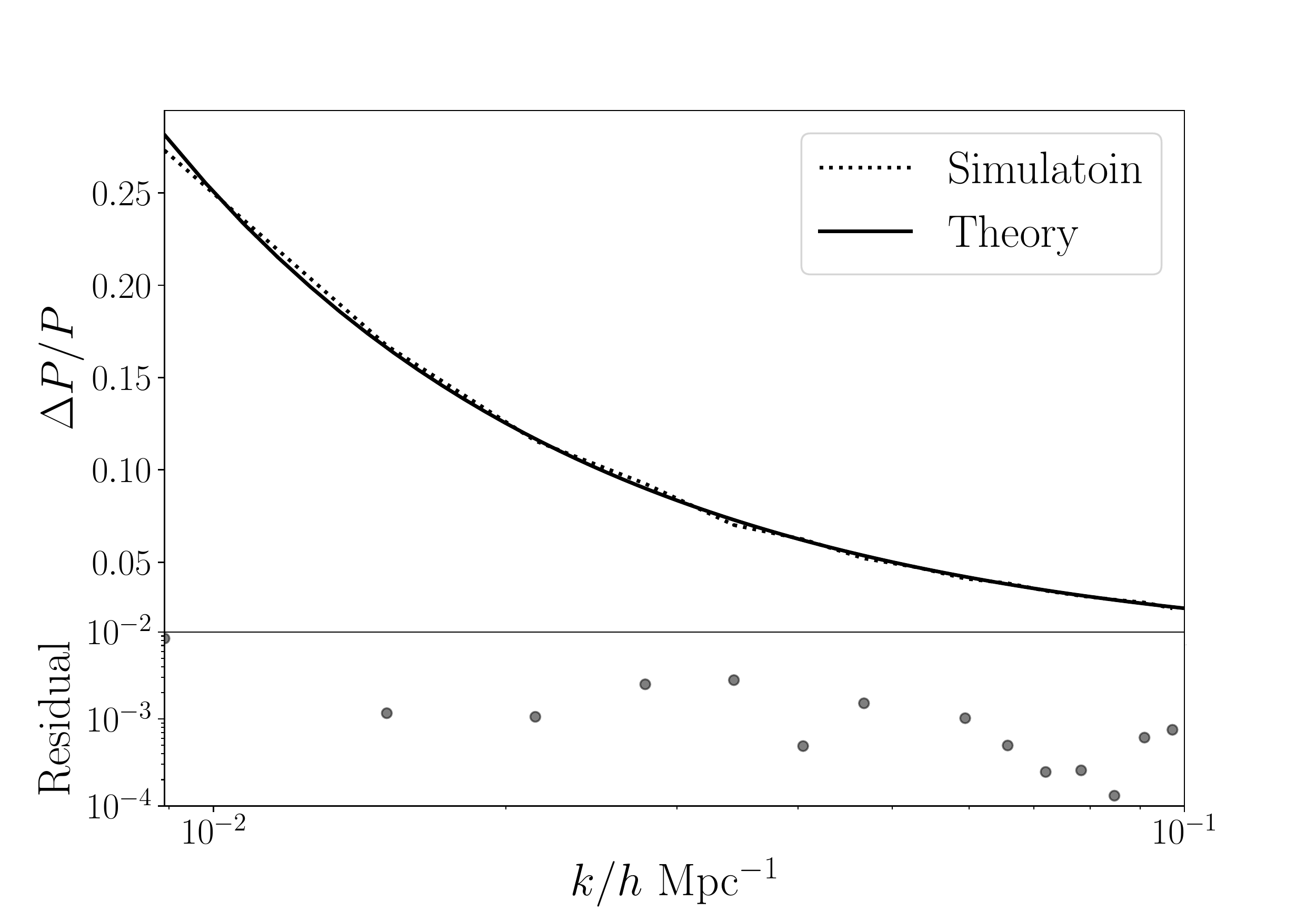}
	\caption{Left panel: the correlation matrix of $P^{(p+s)}_h(K_i)$ defined in Eq.(\ref{eq:cov}); Upper right panel: $\Delta P(k)/\bar{P}(k)$ for power spectrum from simulations (dotted line) and theory prediction (solid line); Lower right panel: corresponding residuals (in absolute value).}
	\label{fig:cov}
\end{figure}

In the left panel of Fig. \ref{fig:cov} we plot the correlation matrix of $P^{(p+s)}_h(K_i)$, defined as
\be
\label{eq:cov}
\frac{C^*_{K_i,K_j}}{\sqrt{C^*_{K_i,K_i}C^*_{K_j,K_j}}},
\ee
where $C^*_{K_i,K_j}$ is the estimated covariance of $P^{(p+s)}_h(K_i)$.
As expected   from  Eq.(\ref{eq:pg}),  skew spectrum and power spectrum are highly coupled at the same $k$-mode. However in the linear-quasi-linear regime of interest here  different $k$ modes are very weakly correlated.

Ref. \cite{hartlap2007your} pointed that the inverse of the maximum-likelihood estimator of the covariance matrix is a biased estimator of the inverse population covariance matrix, and this bias depends on the ratio of the dimensionality  of the matrix $p$ to the number of independent observations $n$. This bias can be corrected by introducing  a Hartlap factor \citep{hartlap2007your},
\be
 C^{-1} = \frac{n-p-2}{n-1} (C^*)^{-1}.
\ee
In our analysis, $n=1000$ and $p=30$, which only leads to a percent level correlation which we include.

 In the right panel of Fig. \ref{fig:cov}, we also show the relative error on the power spectrum $\Delta P_h(k)/\bar{P_h}(k)$ obtained from the simulations  and  the theory prediction,
\be
\Delta P_h(k)/\bar{P_h}(k)=\frac{\sqrt{(2\pi)}}{\sqrt[3]{V}k}
\ee
where $V$ is the simulation volume. From the residuals we can appreciate that this estimation is unbiased.
The errors are large  at large scales due to cosmic variance.

\subsection{Fitting procedure and  error estimate}
We next consider one of the 1000 realisations as our ``mock" Universe to try to constrain  its cosmological parameters by fitting the power spectrum and skew spectrum; {maximum and minimum  scales are: $k_{\min}=0.0089~h\rm Mpc^{-1}$ and $k_{\max}=0.1~h\rm Mpc^{-1}$, respectively.   Here we only consider the tree-level model of the power spectrum since we find that it is sufficient for our purposes  in the range of scales considered (i.e. adding higher order corrections does not change the final results). This can be understood  by considering  non only that one loop corrections are small, but also that the covariance is obtained directly from simulations and we are only interested in the relative errors  (i.e., with/without  skew spectrum).}

We modified the public software $\textsc{CosmoMC}$ \footnote{http://cosmologist.info/cosmomc/} \citep{lewis2002cosmological}, a Markov Chain Monte Carlo (MCMC) code to  calculate the skew spectrum and perform joint Bayesian parameter inference. A simple $\chi^2$ is used for  parameter fitting in our analysis:
\be
\chi^2=\left[\hat P_h^{(p+s)}(K_i)-P^{(p+s)}(K_i)\right]C^{-1}_{K_i,K_j}\left[\hat P_h^{(p+s)}(K_j)-P^{(p+s)}(K_j)\right]^T,
\label{eq:chisq}
\ee
where $\hat P_h^{(p+s)}$ and $P_h^{(p+s)}$ represent the model and the measured spectra.
We recognise that in principle one should use a more appropriate likelihood, the adopted procedure would be correct only if the data vector has a Gaussian distribution which covariance matrix does not depend on the parameters to be estimated. It is well known that the power spectrum does not follow a Gaussian distribution, however the adoption of Eq.(\ref{eq:chisq}) is a good approximation especially for well populated bandpowers see. e.g., \cite{carron2013assumption} and Refs. therein. The distribution for the skew spectrum is certainly non-Gaussian so the adoption of Eq.(\ref{eq:chisq}) is not {\it a priori} justified, but we expect it would  be valid with the central limit theorem.  Here, we  adopt Eq.(\ref{eq:chisq})  as our {\it ansatz}   and we will assess its performance and show it is a sufficiently good approximation below.

Thus, with this in mind,  the best-fit parameters are obtained by finding the minimal of $\chi^2$, and the confidence regions are then defined by the surfaces of constant $\Delta \chi^2=\chi^2-\chi_{\min}^2$, where $\chi_{\min}^2$ is the minimal value of $\chi^2$ and $\Delta \chi^2$ are functions of the number of parameters for the joint confidence levels.

We then repeat the fitting process for the other 999 simulations only to find their best-fit parameter values. By doing this, we can check that  the scatter of recovered parameters among the simulations is consistent with the confidence contours given by the  MCMC-based inference. This is the test that supports our adoption of Eq.(\ref{eq:chisq}).

\section{Results}
\label{sec:result}

\begin{figure}[htb]
	\centering
    \includegraphics[width=0.7\linewidth]{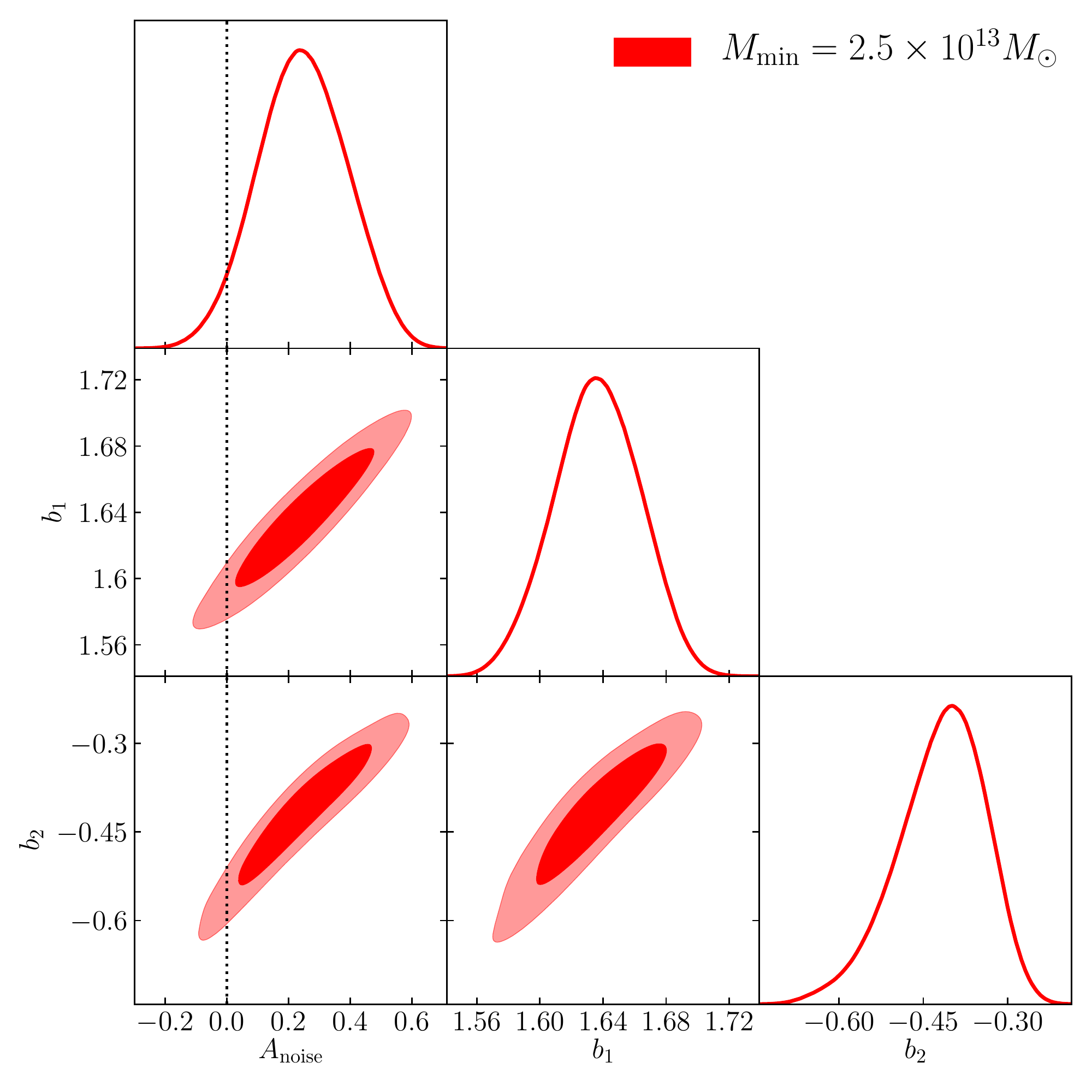}
	\caption{Marginalized two-dimensional distributions (1$\sigma$ and 2$\sigma$ contours) and posterior distributions for $A_{\rm noise}$, $b_1$ and $b_2$. This run is used only to set  $A_{\rm noise}$.}
	\label{fig:Anoise}
\end{figure}

We start by determining the shot noise correction term  and the bias parameters $\{A_{\rm noise}, b_1, b_2\}$  simultaneously  with the fiducial cosmology fixed. Results are shown in  Fig. \ref{fig:Anoise} and Tab. \ref{tab:Anoise}.

We find that $b_1=1.635\pm 0.028$, $b_2=-0.426\pm0.081$ and $A_{\rm noise}=0.22\pm0.15$ (1$\sigma$ C.L.) which  shows a  slight sub-Poisson shot noise. The relationship between $b_1$ and $b_2$ is also consistent with  expectations for the selected halos \citep{lazeyras2016precision, desjacques2018large}.  {We find no difference when adopting $b_{K^2}=0$ or $b_{K^2}=-4/7(b_1-1) $, this is further discussed in appendix \ref{appendix:bK}}.  {To check consistency, we calculate the best-fit results of $b_1$ and $A_{\rm noise}$ using a direct measurement from the ratio of the halo power spectra and the dark matter power spectra at scales $0.0089<k<0.1$ $h$/Mpc. Using 1000 \textsc{Quijote} realizations, the  fitting results of $b_1$ and $A_{\rm noise}$ are $b_1=1.629\pm 0.021$ and $A_{\rm noise} = 0.183\pm0.114$ ($1\sigma$ C.L.), which confirms our adopted values.} {The same analysis repeated in redshift space show fully consistent results, see appendix \ref{appendix:RSD}.}
\begin{table}[h]
	\caption{The best-fit results of $A_{\rm noise}$, $b_1$ and $b_2$ and  their (marginalized,  1$\sigma$) errors. This run is used only to set  $A_{\rm noise}$.}
	\begin{tabular}{c c c}
    \hline
    \hline
    $b_1$&$b_2$&$A_{\rm noise}$\\
    \hline
    $1.635\pm 0.028$ & $-0.426\pm0.081$ & $0.22\pm0.15$\\
    \hline
	\end{tabular}
	\centering
	\label{tab:Anoise}
\end{table}

Now we fix $A_{\rm noise}$ at the best-fit value, and proceed to investigate the potential offered by the combination of power spectrum and skew spectrum. This is motivated by the fact that in a real application the shot noise  and its correction can be determined much more accurately than we can do here by using  for example non-linear scales (where the shot noise dominates).

In this paper we mainly focus on  5 parameters $\{A_s, n_s, f_{\rm NL}^{\rm loc}, b_1, b_2\}$, where $A_s$ and $n_s$ are the  amplitude and spectral index of the primordial spectrum. The other cosmological parameters have been fixed at their fiducial values. It is worth mentioning that the $\textsc{Quijote}$ simulations do not include primordial non-Gaussianity,  hence we   should recover $f_{\rm NL}^{\rm loc}=0$ within  error bars.
While the primordial non-Gaussianity contribution to the  skew spectrum is shown in Eq.(\ref{eq:pris}),  the halo power spectrum can be greatly affected by relatively small values of $f_{\rm NL}^{\rm loc}$ via the non-Gaussian large-scale bias \citep{dalal2008imprints, grossi2009large, wagner2010n,mcdonald2008primordial, matarrese2008effect, sefusatti2009constraining, Dai:2019tjh},
{\be
\frac{\Delta b_1}{b_1-1}=2 f_{\rm NL}^{\rm loc} \frac{\delta_{c}}{M(k,z)W_R(k)} q~,
\ee}
where $\delta_c\simeq 1.686 $ is the threshold for collapse and the correction $q=0.75$  is calibrated  from  $N$-body simulations \citep{grossi2009large}.

\begin{figure}%[htb]
	\centering
    \includegraphics[width=1\linewidth]{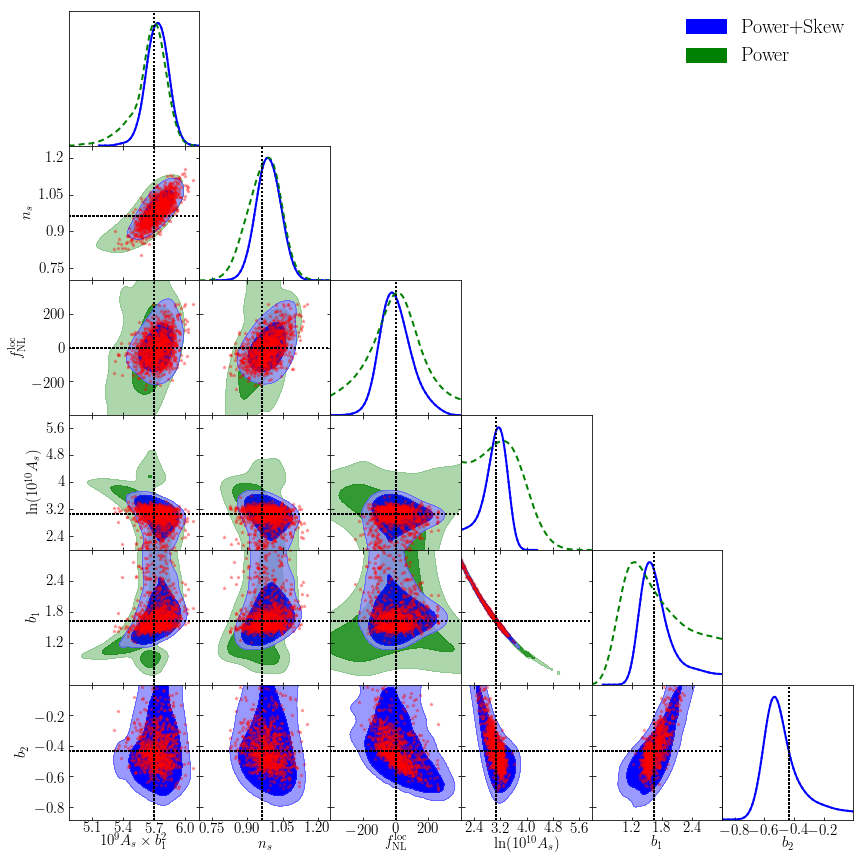}
	\caption{Marginalized two-dimensional distributions (1$\sigma$ and 2$\sigma$ contours) and posterior distributions for $10^9\times b_1^2A_s, n_s, f_{\rm NL}^{\rm loc}, \ln(10^{10}A_s), b_1$ and $b_2$ from power spectrum only (green) and power spectrum together with skew spectrum (blue). The red stars indicate the best-fit points from 1000 simulations using power spectrum and skew spectrum, and black dotted lines are their input  values used in the $\textsc{Quijote}$ simulations.}
	\label{fig:all}
\end{figure}

{In Fig. \ref{fig:all} we show the marginalized 2-D contours for all the parameters using the power spectrum alone and the combination of power spectrum and skew spectrum. We can see $A_{s}, b_1, b_2$  are  highly correlated and show  a very strong degeneracy. Therefore we construct a new variable $10^9\times b_1^2A_s$ to quantify the extra information the skew spectrum can give us. From the results, we find the constraints  are consistent with the fiducial values, indicating that the procedure is not biased. As expected,  adding the skew spectrum results in  tighter constraints. For a more quantitative estimate of the constraining power of the skew spectrum,  we list the best-fit values and their marginalized $1\sigma$ errors in Tab. \ref{tab:all}. The addition of the skew spectrum to the power spectrum yields  a  reduction of the errors  by $31\%, 22\%, 44\%$ for $b_1^2A_s, n_s$ and $f_{\rm NL}^{\rm loc}$ respectively. The constraints on $b_1$ and $A_s$ are unrepresentative since they are strongly degenerate. }

{The figure also shows the scatter of the best-fit results  (power spectrum and skew spectrum combined) of the 1000 realisations is consistent with the confidence contours obtained by  the MCMC-based inference. This indicates that our {\it ansatz} for the likelihood in Eq.(\ref{eq:chisq}) is sufficiently accurate for this application for the joint analysis (perhaps  not unexpectedly it is not too good for poorly constrained parameters  which enter only in the skew spectrum).}
{We find that if $A_{\rm noise}$ is allowed to vary as a nuisance parameter,  along with the other cosmological parameters, the resulting cosmological constraints are weaker,  but the gain arising from adding the skewed spectrum to the power spectrum is greater.  In particular we find that  $1\sigma$  marginalized errors for $b_1 A_s, n_s, f_{\rm NL}^{\rm loc}$ are reduced by $44\%, 39\%, 47\%$ by including the skew spectrum.
Details are in appendix \ref{appendix:Anoise}.}

{We repeated the analysis on redshift space  for the power spectrum and skew spectrum monopole and find that by combining the skew spectrum to the power spectrum
the 1-$\sigma$ marginalized errors $b^2A_s$, $n_s$ and $f^{\rm loc}_{\rm NL}$ are reduced by 41\%, 26\%, 39\% respectively for  fixed $A_{\rm noise}$. The details are  reported in Appendix~\ref{appendix:RSD}.}

The error reduction provided by the inclusion of the skew spectrum can be understood as follows.
 Eq.(\ref{eq:pg}) and Eq.(\ref{eq:mathf}) indicate that the  (tracer) skew spectrum can be seen as a (tracer) power spectrum  modulated by a scale-dependent function, whose amplitude and scale dependence depend on the  key parameters  with a scaling that is different from the power spectrum dependence.  Hence the additional information enclosed in  $\mathcal{F}(k)$ can be  used to  reduce degeneracies  among parameters that are present at the level of the power spectrum.

\begin{table}[htb]
	\caption{The best-fit results of $10^9\times b_1^2A_s, n_s$ and $f_{\rm NL}^{\rm loc}$, together with their  marginalized  1$\sigma$ errors.}
	\begin{tabular}{c c c}
    \hline
    \hline
    Parameters&Power spectrum&Power + skew spectrum\\
    \hline
    $10^9\times b_1^2A_s$ & $5.661\pm0.163$ & $5.714\pm0.112$\\[3pt]
    $n_s$ & $0.972\pm0.068$ & $0.983\pm0.054$\\[3pt]
    $f_{\rm NL}^{\rm loc}$ & $-2.9\pm167.8$ & $-8.9\pm94.3$\\[3pt]
    $\ln(10^{10}A_s)$ & $2.37^{+0.84}_{-1.32}$ & $3.31_{-0.77}^{+0.29}$\\[3pt]
    $b_1$ & $2.29^{+0.71}_{-1.48}$ & $1.45_{-0.31}^{+0.69}$\\[3pt]
    $b_2$ & --- & $-0.51^{+0.52}_{-0.41}$\\[3pt]
    \hline
	\end{tabular}
	\centering
	\label{tab:all}
\end{table}

%In Fig. \ref{fig:b12} we show the  marginalised, two-dimensional distributions of $b_1$, $b_2$ and $\ln(10^{10}A_s)$ using power spectrum and skew spectrum;  the degeneracy between these parameters is very clear (underlying  values are still recovered within the errors).

%\begin{figure}[htb]
%	\centering
%    \includegraphics[width=0.8\linewidth]{deng.pdf}
%	\caption{Marginalized two-dimensional distributions of $b_1$, $b_2$ and $\ln(10^{10}A_s)$ using the combination of power spectrum and skew spectrum.}
%	\label{fig:b12}
%\end{figure}

\section{Conclusions and discussion}
\label{sec:con}

In this paper, we have considered a relatively unexplored statistic, the  skew spectrum,  which is estimated using the cross spectrum of the  squared density  field $\delta^2(\boldsymbol x)$ with the field  $\delta(\boldsymbol x)$ itself. Computationally, evaluation of skew spectrum is equivalent (in terms of speed and complications) to a power spectrum estimation, but the skew spectrum contains 3-point clustering (bispectrum) information.  While the use of the full bispectrum provides optimal constraints, and it is the correct approach to access all the information enclosed in the three-point function,  its practical implementation is challenging. This  has motivated the search for alternative statistics that can capture partial information but a much reduced cost (and improved speed). One of them is the integrated bispectrum \cite{chiang2014position}, the skew spectrum is another, complementary,  alternative.

We have derived  the general form of skew spectrum and then considered three main contributions: the primordial non-Gaussianity, gravitational instability and galaxy (halo) bias {both in real space and redshift space}. Finally we expressed this specific skew spectrum as a function of a scale-dependent function, $\mathcal{F}(k)$, times power spectrum. Because of the non-linear nature of  $\mathcal{F}(k)$, and its peculiar dependence on key parameters such as  the bias and non-Gaussianity parameters, the skew spectrum offers an extra handle on these. We have built on the works of Refs.\cite{pratten2012non, schmittfull2015near, dizgah2019capturing} who also consider the skew spectrum for primordial non-Gaussianity, bias and gravitational evolution. {However here we address a different issue: we ask what  can be the added value of combining  skew spectrum and power spectrum in a joint analysis.  This has not been addressed before in the literature, but it is a question worth asking before deciding whether to proceed to measure the skew spectrum from real galaxy surveys. Usually these type of analyses are done via a Fisher matrix approach.  Here however,  because of  the high covariance between power spectrum and Skew spectrum, and the complex properties of the skew-spectrum covariance, we  resort to numerically computed covariance from a suite of  1000 state-of-the art simulations.}

We  have compared  both the performance  of our modelling for the skew spectrum and simulation results by resorting to the $\textsc{Quijote}$ suite.  We  find that  our analytic modelling of the skew spectrum reproduce the simulations well if the cosmological density (or halo) field is  smoothed on linear or quasi-linear scales. The addition of the skew spectrum  to the power spectrum provides a reduction of  $31\%, 22\%, 44\%$ on the error of the parameters $ b_1^2A_s, n_s$ and $f_{\rm NL}^{\rm loc}$ respectively. { this gains also hold in redshift space when considering only the monopole.}  However, by limiting the analysis to  the { linear  scales considered here (we adopt a top hat smoothing filter of $20$ Mpc$/h$)},  the skew spectrum only lift very partially the degeneracy between $A_s$ and the linear and quadratic bias parameters.  Nevertheless  a reduction of $44\%$ on the error of $f_{\rm NL}^{\rm loc}$ is interesting as it would correspond to roughly doubling the survey volume if one  were  to use the power spectrum only (in the same $k$-range). It is possible that with a more sophisticated modelling of  the gravitational instability kernel, the analysis could be pushed to higher $k$ further lifting the remaining degeneracies.

Because of its simplicity,  the use of the skew spectrum in a standard pipeline for analysis of galaxy surveys could  offer a powerful  and fast cross check  for  possible systematics errors in a joint power spectrum+bispectrum analysis.
Moreover, we envision that a statistics like the skew spectrum could be used,  instead of the bispectrum in a practical application to a galaxy survey along the power spectrum,  if one is only interested in reducing the errors on the linear and quadratic bias parameters. In fact Ref. \cite{schmittfull2015near}  shows that for initially gaussian fields,
this statistics is (near) optimal, in the sense that  if used with inverse variance weighting, it captures virtually all the information present in the angle-independent
part of the bispectrum, and therefore in the combination $b_1^2 b_2$. Here we do not use the optimal weighting,  but the resulting statistics, while sub-optimal,  is still unbiased.   We conclude by acknowledging that while only the auto skew spectrum was considered here, in the present era of multi-tracers cosmology, the  (cross) skew spectrum can be a much richer quantity. For example, given two  tracers, $i$ and $j$  of the same (density) field, one could form   4 cross skew spectra $\delta_i^2$$\times$ $\delta_j$, $\delta_j^2 \times \delta_i$, $\delta_i\delta_j \times \delta_j$, $\delta_i\delta_j \times \delta_i$, compared to one cross power spectrum $P_{ij}$. We envision that the combination of the (auto+cross) skew spectra  to the (auto+cross) power spectra could be very synergetic, both in terms of reducing error bars on cosmological parameters and in helping to control  possible systematic errors in the measurement and/or its interpretation.  We leave this exploration to future work.

\section*{Acknowledgements}
JPD thanks the ICCUB (Institut de Ciencies del Cosmos, University de Barcelona) for hospitality.
LV acknowledges support of European UnionÕs Horizon 2020 research and innovation programme ERC (BePreSySe, grant agreement 725327). Funding for this work was partially provided by the Spanish Ministerio de Ciencia y Innovation y Universidades  under project PGC2018-098866-B-I00. JQX acknowledges support of the National Science Foundation of China under grants No. U1931202, 11633001, and 11690023; the National Key R\&D Program of
China No. 2017YFA0402600; the National Youth Thousand
Talents Program and the Fundamental Research Funds for the
Central Universities, grant No. 2017EYT01.
We acknowledge the use of the Quijote simulations https://github.com/franciscovillaescusa/Quijote-simulations.

\appendix

\section{Effect of  $b_{K^2}$}
\label{appendix:bK}
\begin{figure}[ht]
	\centering
    \includegraphics[width=0.6\linewidth]{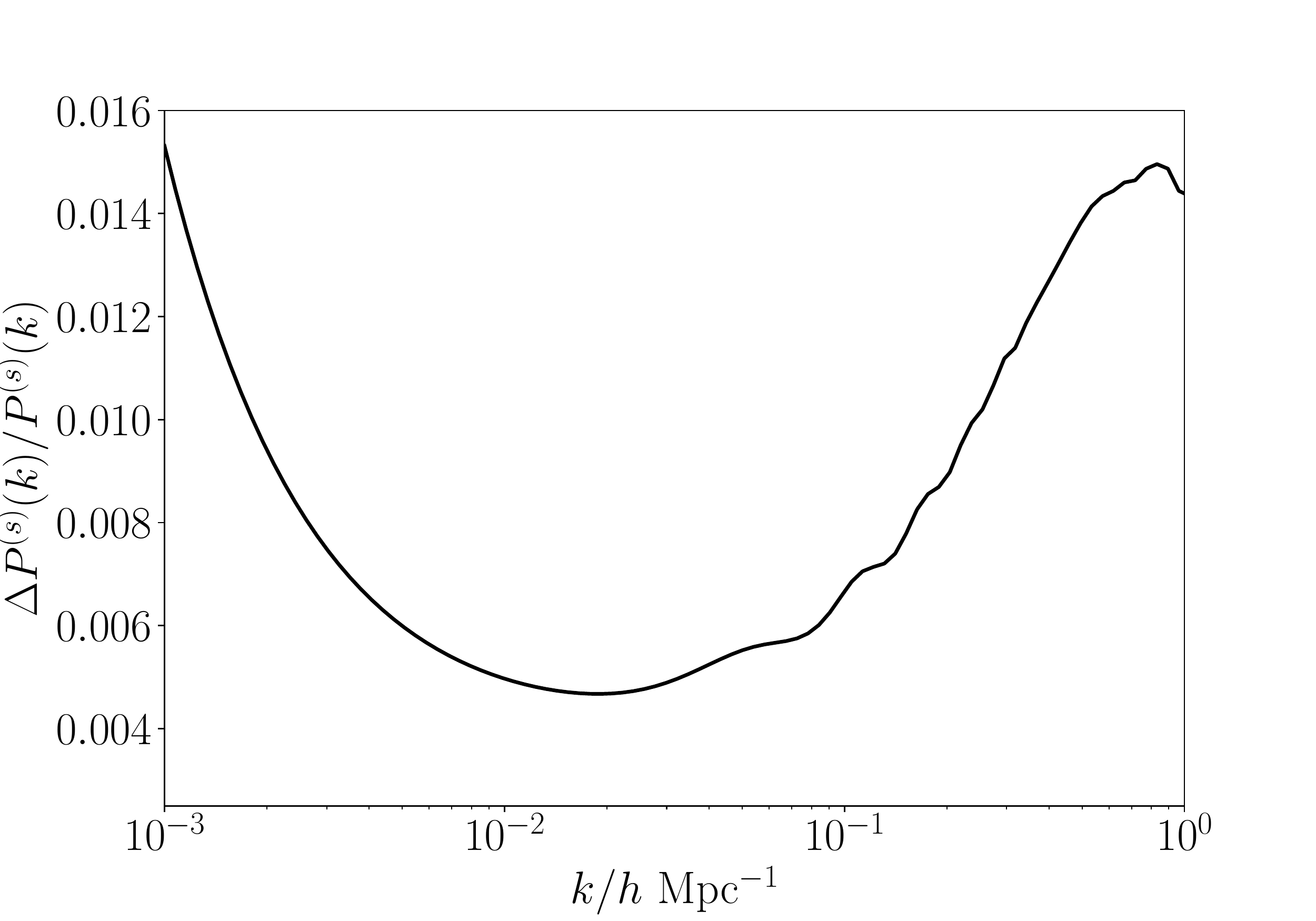}
	\caption{The relative difference in the skew spectrum  induced by the tidal bias with $b_{K^2}=-4/7(b_1-1)$ where we fix $b_1=1.62, b_2=-0.43, f_{\rm NL}^{\rm loc}=0$ and the power spectrum adopted is a Planck $\Lambda$CDM model. Note that the scales adopted in this work  are $k<0.1 h/Mpc$.}
	\label{fig:dbk2}
\end{figure}

\begin{figure}[htb]
	\centering
    \includegraphics[width=1\linewidth]{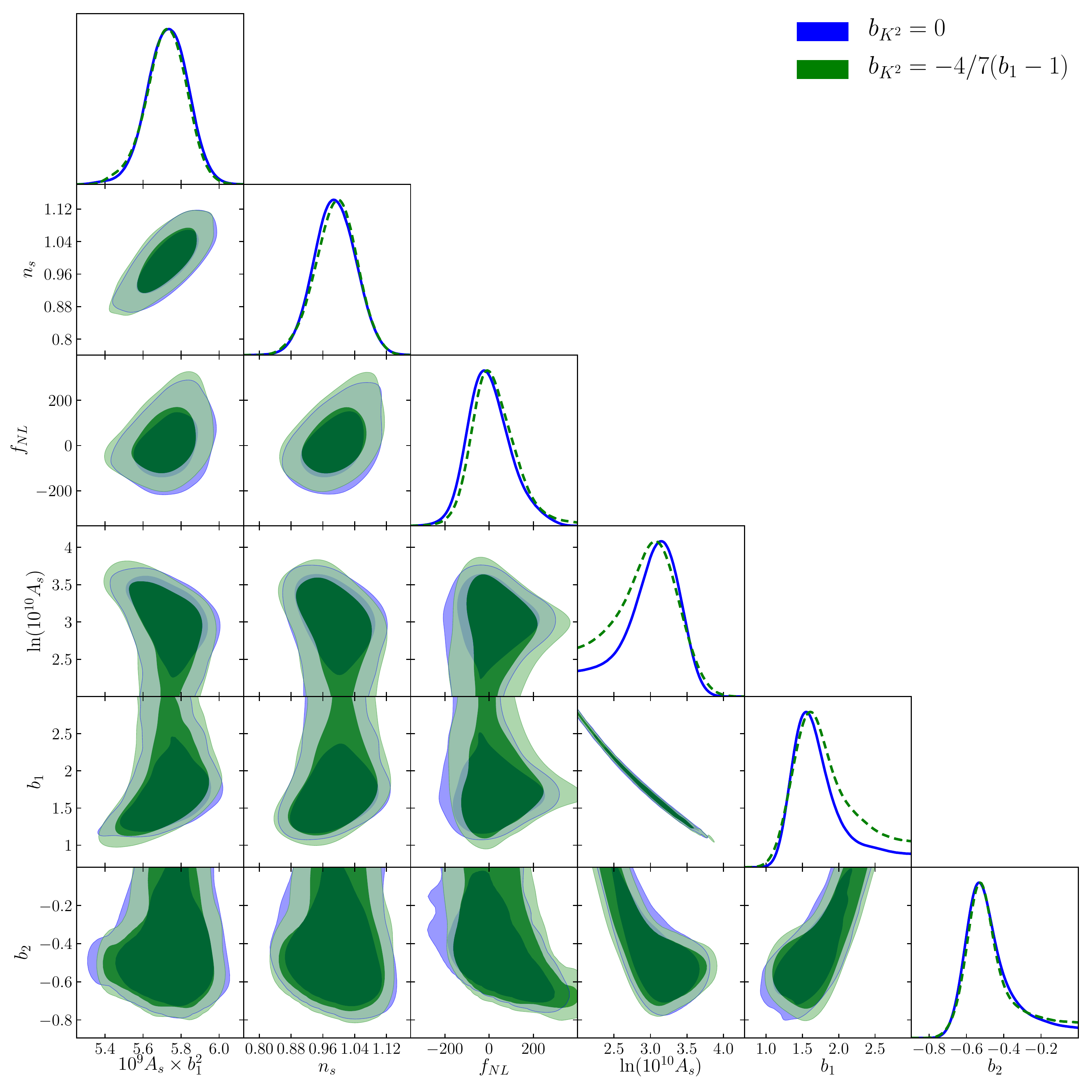}
	\caption{Effect of the adopted value for $b_{K^2}$. Marginalized two-dimensional distributions and posterior distributions for all the parameters of interest. We use the combination of power spectrum and skew spectrum and consider $b_{K^2}=-4/7(b_1-1)$ (green) and $b_{K^2}=0$ (blue, results in  the main text.). This shows that our results are insensitive to the specific value of  $b_{K^2}$ in this range.}
	\label{fig:cbk2}
\end{figure}

{Here we show that the exact value adopted for   $b_{K^2}$ does not affect our conclusions. Fig. \ref{fig:dbk2} shows that the effect on the skew spectrum (in real space) is at the 1\% level. We compare results  obtained for two  values:  $b_{K^2}=0$ (no shear bias, as if bias was local in Eulerian space) and  $b_{K^2}=-4/7(b_1-1)$ (bias is local in Lagrangian space. This is shown in Fig.~\ref{fig:cbk2} for real space.}

\newpage

\section{Effect of  treating $A_{\rm noise}$ as a free parameter in the joint fit.}
\label{appendix:Anoise}
\begin{figure}[htb]
	\centering
    \includegraphics[width=1\linewidth]{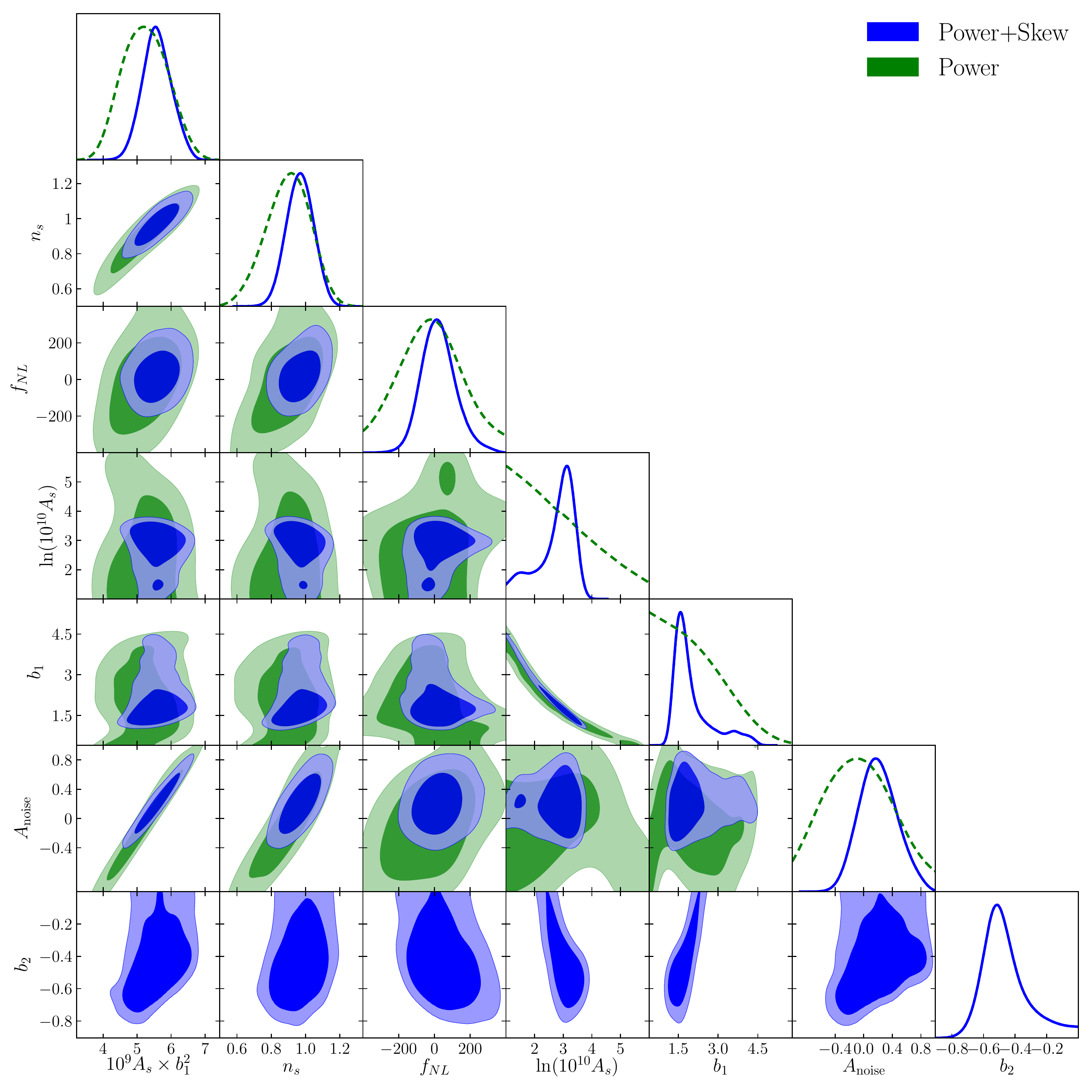}
	\caption{This is the analogous of Fig.~\ref{fig:all} in the main text. Parameters constraints  in real space, but  treating $A_{\rm noise}$ as a free parameter. The conclusions to not change significantly but the errors are are larger due to the degeneracies with $A_{\rm noise}$ (especially $n_s$ and $b_1^2A_s$).}
\label{fig:rpan}
\end{figure}

\begin{figure}[htb]
	\centering
    \includegraphics[width=1\linewidth]{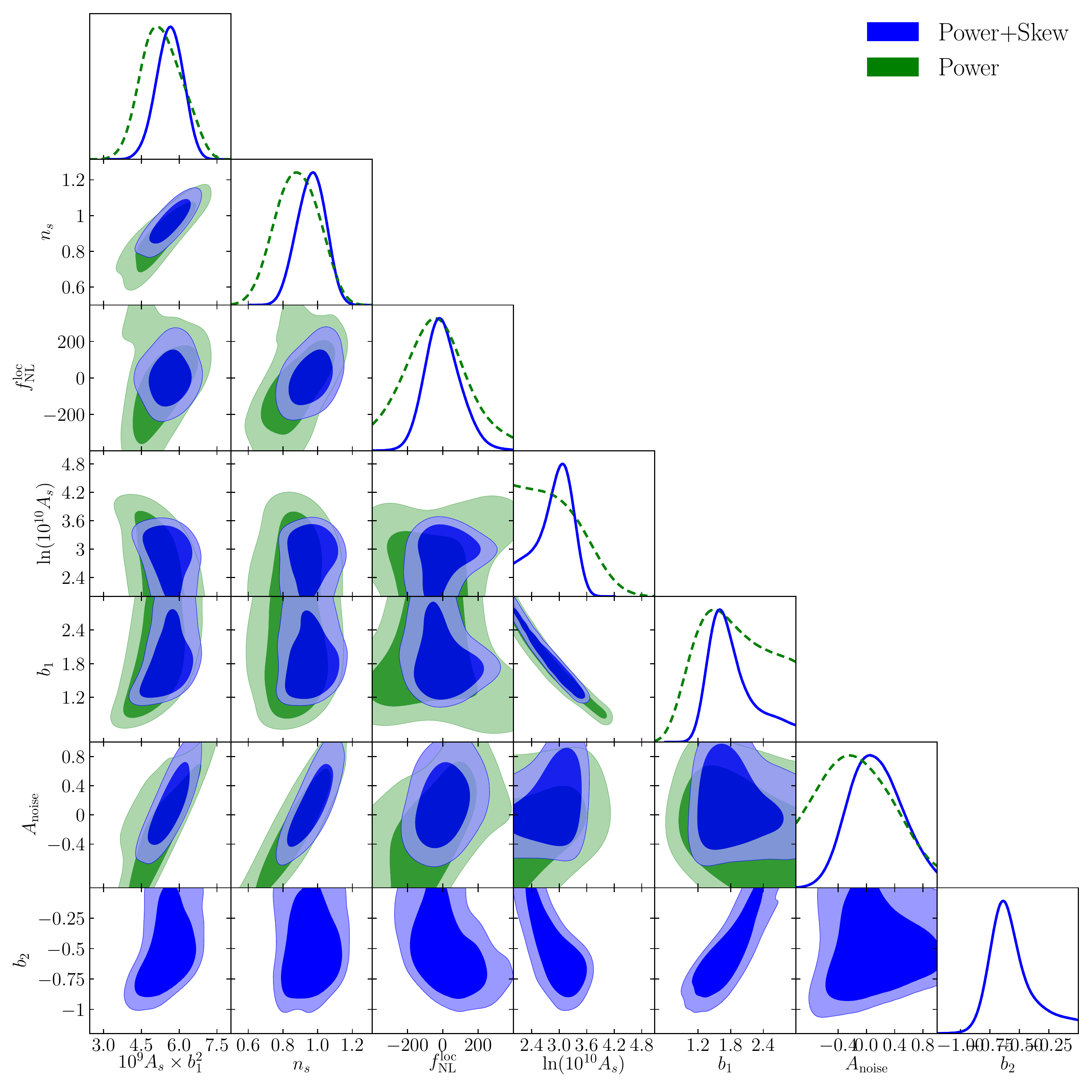}
	\caption{Same as Fig.~\ref{fig:rpan} but in redshift space.}
	\label{fig:RSD2}
\end{figure}

{Here we show  that, allowing  $A_{\rm noise}$ to float along the other cosmological parameters in the joint fit does not invalidate our findings. On the contrary it yields slightly better forecasted improvement when adding the skew power spectrum. Fig.~\ref{fig:rpan} is the analogous to Fig.~\ref{fig:all} in the main text in real space, Fig.\ref{fig:RSD2} is in redshift space.}

\newpage

\section{Redshift space analysis}
\label{appendix:RSD}
{ While we have presented real space results in the main text  and discussed how the redshift space analysis yields very similar conclusions we report the details here.}

\begin{figure}[htb]
	\centering
    \includegraphics[width=1\linewidth]{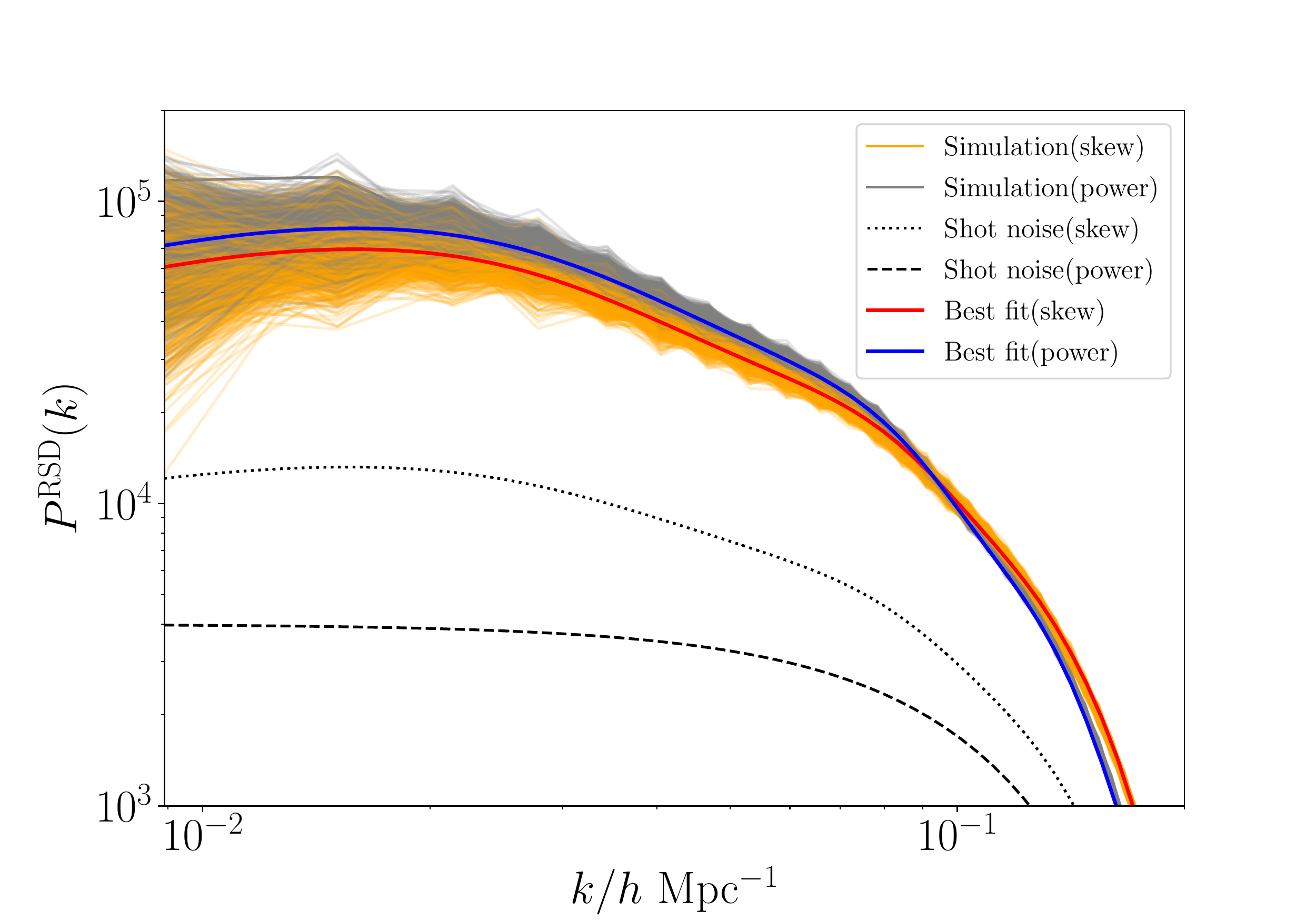}
	\caption{Same as Fig. \ref{fig:shot} in  the main text but in redshift space. The best-fit bias parameters and noise correction are: $b_1=1.61, b_2=-0.62, A_{\rm noise}=0.16$}
	\label{fig:sm20}
\end{figure}

\begin{figure}[htb]
	\centering
    \includegraphics[width=1\linewidth]{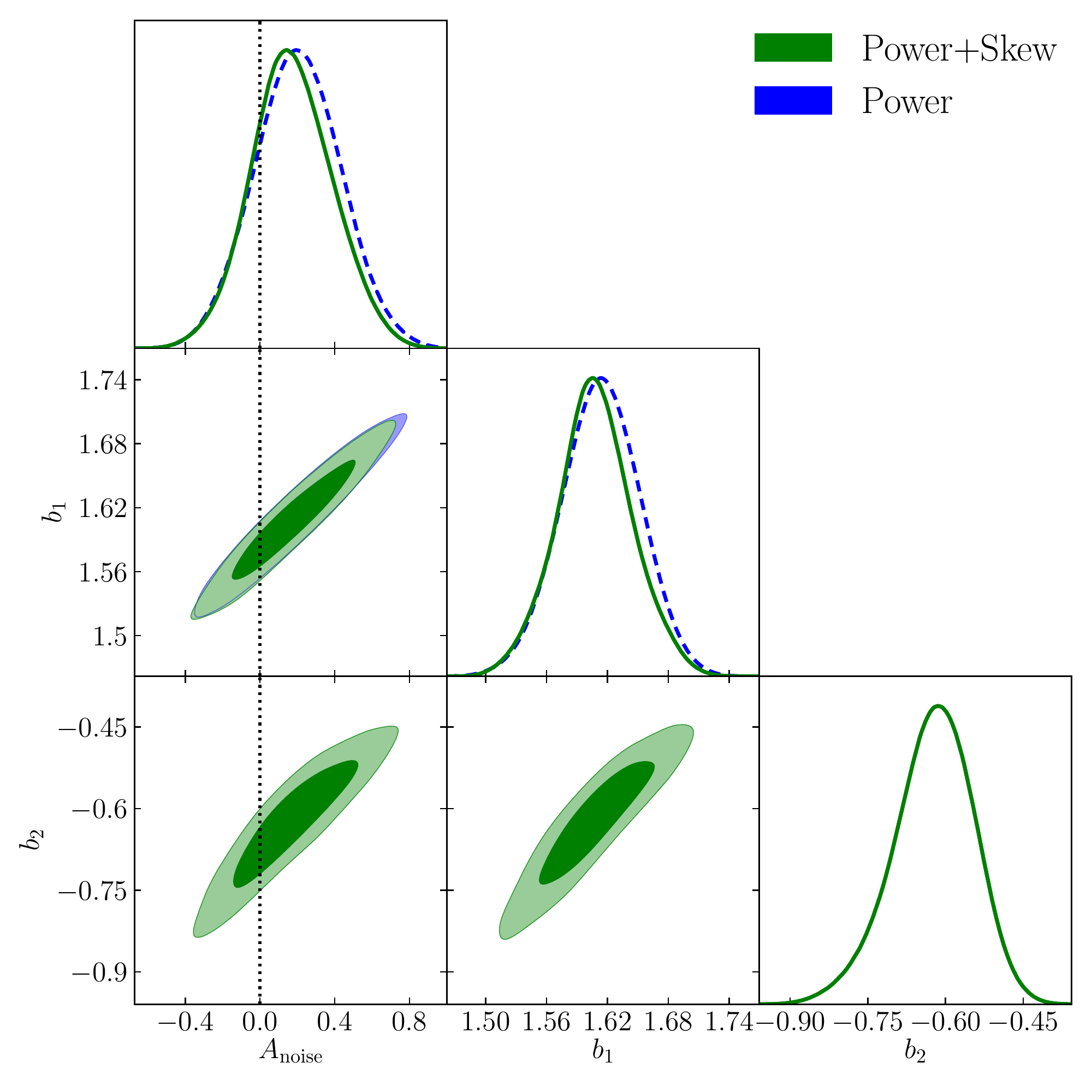}
	\caption{Same as Fig.~\ref{fig:Anoise} in   the main text  but in redshift space. When adopting the combination of power spectrum and skew spectrum, the marginalized $1\sigma$ constraints  are: $b_1=1.606\pm0.036, b_2=-0.619\pm0.077, A_{\rm noise}=0.164\pm0.218$, fully consistent with the real space analysis.}
\label{fig:RSD1}	
\end{figure}

{In Fig.~\ref{fig:sm20} and  Fig.~\ref{fig:RSD1}  we show the redshift-space fits to the skew spectrum and for the  bias parameters and $A_{\rm noise}$, these figures correspond to Figs.~ \ref{fig:shot} and \ref{fig:Anoise} of the main text respectively.  Here we use only the monopole signal. While the anisotropic signal in  the power spectrum has been extensively used in the literature this is not the case for the bispectrum. For this reason for this initial investigation we limit ourselves to the monopole for both statistics.  The resulting constraints on cosmological parameters of interest (also varying $A_{\rm noise}$) are shown in Fig. \ref{fig:RSD2}. In summary in redshift space adding skew spectrum
to power spectrum the $1-\sigma$ marginalized errors for parameters $b_2A_s$, $n_s$ and $f_{\rm NL}^{\rm loc}$ are reduced by 41\%, 26\%, 39\% respectively.}

%\bibliography{ref}{}
%\bibliographystyle{JHEP}

\providecommand{\href}[2]{#2}\begingroup\raggedright\endgroup

\end{document}